\documentclass[11pt]{article}
\usepackage{mathrsfs, amssymb, amsfonts, amsthm, amsmath, amsbsy, bbm, bm, mathtools, dsfont}
\usepackage[hidelinks]{hyperref}
\usepackage{ytableau}
\usepackage{comment}
\usepackage{xcolor}
\usepackage{autonum}
\usepackage{graphicx}
\usepackage{enumerate}
\usepackage{natbib}
\usepackage{url} 
\usepackage{xr}
\usepackage[linesnumbered,ruled, vlined]{algorithm2e}

\renewcommand{\baselinestretch}{1}
\allowdisplaybreaks 

\newtheorem{definition}{Definition}
\newtheorem{proposition}{Proposition}
\newtheorem{theorem}{Theorem}
\newtheorem{corollary}{Corollary}
\newtheorem{remark}{Remark}

\newcommand{\len}{\ell}
\def \N {\mathbb{N}}
\def \PP {\mathrm{Pr}}
\def \P {\mathcal{P}}
\def \d {{\rm d}}
\def \H {\mathcal{H}}
\def \F {\mathcal{F}}
\def \Z {\mathbb{Z}}
\def \T {\mathcal{T}}
\def \S {\mathcal{S}}
\def \M {\mathcal{M}}
\def \E {\mathbb{E}}
\def \B {\mathcal{B}}
\def \dualtr {p^{\downarrow}}

\def \ind {{\mathds{1}}}	
\def \iid {\overset{\text{iid}}{\sim}}
\def \simplex {\overline\nabla}
\def \simplexone {\nabla}
\def \tppd {two-parameter Poisson--Dirichlet$\ $}
\def \PD {\mathrm{PD}_{\alpha, \theta}}
\def \pd {\mathrm{PD}}
\def \pr {\Pr_{\alpha,\theta}}
\def \prp {\mathscr{P}_{\alpha,\theta}}
\def \sp {\preccurlyeq}
\def \psf{\mathrm{PSF}_{\alpha, \theta}}
\def \coag{\mathrm{coag}}
\def \Coag{\mathrm{Coag}_{\alpha, \theta}}
\def \csp {C_{\simplexone}([0,\infty))}
\def \CRP{\mathrm{CRP}}
\def \paint{\text{Part}}
\def \dual{\mathrm{Dual}}

\title{\bf \vspace{-15mm}
Exact inference via quasi-conjugacy in two-parameter Poisson--Dirichlet hidden Markov models}

\author{
{\sc Marco Dalla Pria}\\
\emph{ESOMAS Department, University of Torino}
\and
{\sc Matteo Ruggiero}\\
\emph{Stern School of Business, NYU Abu Dhabi}
\and
{\sc Dario Span\`o}\\
\emph{Department of Statistics, University of Warwick}
}


\begin{document}

\maketitle

\def\spacingset#1{\renewcommand{\baselinestretch}%
{#1}\small\normalsize} \spacingset{1}


\begin{abstract}
We introduce a nonparametric model for inferring time-evolving, unobserved probability distributions from discrete-time data consisting of unlabelled partitions. The latent process is a two-parameter Poisson--Dirichlet diffusion, and observations arise via exchangeable sampling. Applications include social and genetic data where only aggregate clustering summaries are observed.
To address the intractable likelihood, we develop a tractable inferential framework that avoids label enumeration and direct simulation of the latent state. We exploit a duality between the diffusion and a pure-death process on partitions, together with coagulation operators that encode the effect of new data. These yield closed-form, recursive updates for forward and backward inference.
We compute exact posterior distributions of the latent state at arbitrary times and predictive distributions of future or interpolated partitions. This enables online and offline inference and forecasting with full uncertainty quantification, bypassing MCMC and sequential Monte Carlo. Compared to particle filtering, our method achieves higher accuracy, lower variance, and substantial computational gains. We illustrate the methodology with synthetic experiments and a social network application, recovering interpretable patterns in time-varying heterozygosity.
\end{abstract}

\noindent%

\tableofcontents

\spacingset{1.1} 


\section{Introduction}
\label{sec:intro}

We study inference for a latent, continuous-time process of random probability measures observed only through partial summaries. The goal is to recover the law of a time-evolving signal \(X(t)\), which is an infinite-dimensional vector of decreasing relative frequencies, from discrete-time observations given as \emph{unlabelled partitions}.

Such problems arise when the latent structure has unknown or unbounded complexity and only aggregated data are observed. Examples include social networks, where one records sizes of connected components without identities, and biological applications, where species labels are absent and only relative abundances are measured \citep{bunge1993estimating, holmes2012dirichlet, pritchard2000inference}. A defining feature is that observations encode frequencies, not labels.

A natural framework is the \emph{hidden Markov model} (HMM), with latent states evolving via a Markov transition and observations conditionally independent given those states \citep{cappe2005springer}. Classical cases admit exact filters---Kalman, Baum--Welch, Wonham \citep{kalman1960, baum1966statistical, wonham1965applications}---and later work extends to certain diffusions \citep{chaleyat2006computable, chaleyat2009filtering, papas_ruggiero}. In more complex settings, inference typically relies on sequential Monte Carlo (SMC) \citep{chopin2020introduction}.

Here we address a more challenging scenario: the unobserved process $X(t)$ is an \emph{infinite-dimensional} vector of probabilities representing latent type frequencies. The observations $Y(t_k)$, however, are \emph{unlabelled partitions}: they describe how a sample drawn from $X(t_k)$ is clustered, but do not retain information about which probability component of $X(t_k)$ corresponds to each observed group.
The inferential goal is to compute filtering distributions \(p(x_{t_k} |  y_{t_0},\ldots,y_{t_k})\) and related smoothing or predictive laws, which requires structural and computational tools beyond those available for classical HMMs.

Prior work on infinite-dimensional HMMs includes discrete-time models based on Dirichlet and Pitman--Yor processes \citep{beal2001infinite, gael2008infinite, chatzis2010infinite, caron2017generalized} and filters for Fleming--Viot diffusions \citep{papaspiliopoulos2016conjugacy, ascolani2023smoothing}. Here we pursue a continuous-time formulation with latent dynamics given by a two-parameter Poisson--Dirichlet (PD) diffusion \citep{petrov2009, walker2009countable, feng2010some}. This class, which arises in coalescent theory and population genetics \citep{costantini2017wright, griffiths2024dual}, captures power-law clustering but remains largely unexplored in the context of Bayesian nonparametrics.

We therefore propose a new class of infinite-dimensional HMMs in which a PD diffusion governs the latent dynamics while observations take the form of unlabelled partitions. This formulation separates latent evolution from combinatorial observation structure, allowing inference without label tracking and making the model well suited to clustering, social networks, and population dynamics (see Section~\ref{sec:partition-structures}).

\begin{figure}[t!]
    \centering
    \includegraphics[width=.9\linewidth]{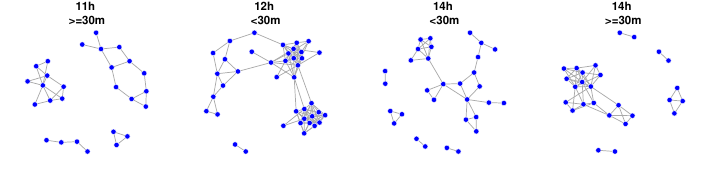}
\caption{\footnotesize Unlabelled partitions obtained over four intervals from the \textsc{Infectious} dataset.}
    \label{fig:infectious}
\end{figure}

A motivating example is the \textsc{Infectious} dataset \citep{isella2011s}, which records time-varying contact networks via wearable sensors. Grouping individuals by connected components yields a sequence of unlabelled partitions, as shown in Figure~\ref{fig:infectious}. We revisit this dataset in Section~\ref{sec: application}.

Existing inference methods fail in this regime. SMC requires simulating the diffusion, whose transition law lacks closed form and must be truncated in practice, introducing bias, while MCMC schemes for Dirichlet or Pitman--Yor mixtures rely on labelled structures that are unavailable in our setting. Formally, the likelihood of partitions combines an unlabelled likelihood that sums over latent label assignments with an intractable diffusion transition density (see Section~\ref{subsec:likelihood}). Both remain computationally prohibitive, and truncation introduces hard-to-quantify bias, motivating a structural alternative.

We exploit a dual representation of the Poisson--Dirichlet diffusion  as a pure-death process on partitions \citep{griffiths2024dual}. This dual evolves backward in time and tracks sufficient statistics of the signal. Combined with algebraic operations on partitions (\emph{coagulations}), it yields recursive, closed-form updates that enable exact Bayesian inference. In particular, we compute posterior distributions of the latent state at any observation time, smoothing distributions at arbitrary intermediate times, and predictive distributions for future or interpolated partitions, all with full uncertainty quantification.

Our method is a nonparametric analogue of the Baum--Welch filter: in finite-state HMMs, Baum--Welch exploits latent structure to enable exact inference; here we exploit the finite combinatorics of the dual pure-death process to obtain structurally exact filters and smoothers for a continuous-time, infinite-dimensional model. When scalability is needed, the method can be paired with targeted Monte Carlo approximations that preserve the exact update structure while avoiding the inefficiencies of generic SMC or MCMC.

\paragraph{Paper outline.} Section~\ref{sec:preliminaries} develops novel results for partitions and the associated likelihood. 
Section~\ref{sec:filtering} presents the inferential framework and the main results. Section~\ref{sec: illustration} reports numerical and real-data illustrations, including a comparison with bootstrap particle filtering. Proofs, algorithms, and further details are given in the Supplement.

\paragraph{Summary of Notation.}
\label{sec:notation} 
For ease of the reader, we collect here some notation used throughout:

\begin{itemize}\topsep0em
    \itemsep0em
    \item \textbf{Sets:} $\N$ is the set of natural numbers; $\P_n$ is the set of unlabelled partitions of $n$, and $\P = \bigcup_{n \ge 1} \P_n$.
    \item \textbf{Partitions:} $\pi = (\pi_1, \dots, \pi_{\ell}) \in \P_{n}$ denotes a partition with $|\pi| = \sum \pi_i = n$ and length $\ell$; $\coag(\omega, \gamma) \subseteq \P_{m+n}$ is the set of coagulations $\mu$ of $\omega \in \P_{m}$ and $\gamma \in \P_{n}$, with combinatorial weights $\H(\omega, \gamma | \mu) \in [0, 1]$.
    \item \textbf{Distributions:} $\PD$ is the Poisson--Dirichlet distribution; $\psf$ is the Ewens--Pitman sampling formula.
    \item \textbf{Models:} $X(t)$ is the latent PD diffusion; $\Pi^{k} := \Pi(t_k)$ is the observed unlabelled partition at time $t_k$; $P(\pi | x)$ is the unlabelled likelihood of $\pi$ given state $x$.
\end{itemize}

 
\section{Partition structures and likelihood representation}
\label{sec:preliminaries}

This section introduces the combinatorial and probabilistic ingredients underlying inference for the two-parameter Poisson--Dirichlet HMM.
We proceed in two conceptually distinct steps.
First, we review the \emph{unconditional} Ewens--Pitman model for random partitions, which describes the law of an exchangeable partition arising from sampling without conditioning on a latent state.
Within this framework, we introduce a combinatorial operation on partitions, called \emph{coagulation}, and establish product expansions associated with the Ewens--Pitman sampling formula.
These identities characterize how independent partition-valued observations combine and yield finite mixture representations.

We then turn to the \emph{conditional} setting relevant for hidden Markov models, where partitions arise from sampling given a latent state $X(t)$ evolving in time.
This leads to an unlabelled likelihood for partitions conditional on $X(t)$, and to analogous product expansions that induce dependence between observations over time.
Together, these results form the basis for the filtering and smoothing methods developed in Sections~\ref{sec:filtering}--\ref{sec: illustration}; full proofs and technical details are deferred to Section~\ref{SMsec:preliminaries}.


\subsection{The Ewens--Pitman sampling formula}
\label{sec:partition-structures}

\begin{figure}[t!]
    \centering
    \includegraphics[width=.63\textwidth]{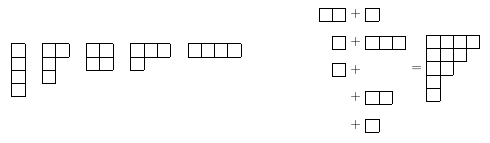}
\caption{\footnotesize 
Left: $\P_4 = \{ (1, 1, 1, 1), (2, 1, 1), (2, 2), (3, 1), (4)\}$ represented via \emph{Young diagrams}. 
Right: an indexed coagulation of \(\omega = (2,1,1)\) and \(\gamma = (3,2,1,1)\), yielding $\mu=(4,3,2,1,1)$.}
\label{fig:diagrams}
\end{figure}

Let $\pi = (\pi_1, \dots, \pi_{\len})$ denote a partition of $n$ of \emph{length} $\len$, i.e.,  with $\len$ non-increasing integer parts such that $|\pi| := \sum_{i=1}^{\len} \pi_i = n$.
We denote the set of all partitions of $n$ by $\mathcal{P}_n$ (see Fig.~\ref{fig:diagrams} (left) for an example), and let $\mathcal{P} = \bigcup_{n \ge 1} \mathcal{P}_n$ be the set of all partitions.

A widely used distribution for random partitions is the \emph{Ewens--Pitman sampling formula}: it assigns to any particular $\pi = (\pi_{1}, \dots, \pi_{\len}) \in \P$ the probability
\begin{equation}\label{PSF}
\psf(\pi) :=  C(\pi) \frac{\prod_{i = 0}^{\len - 1} (\theta + i \alpha ) }{\theta_{(|\pi|)}} \prod_{i = 1}^{\len} (1 - \alpha)_{(\pi_{i} - 1)},
\end{equation}
with $C(\pi)$ a combinatorial coefficient (see \eqref{psf explicit}), $\alpha \in [0,1)$ and $\theta > -\alpha$. 
The factor multiplying $C(\pi)$ in \eqref{PSF} is the \emph{exchangeable partition probability function} (EPPF) of the two-parameter Poisson--Dirichlet distribution \citep{pitman1995exchangeable}, while the coefficient $C(\pi)$ accounts for the number of set partitions having block-size sequence $\pi$.
The associated predictive rules correspond to a \emph{Chinese restaurant process} (CRP) with parameters $(\alpha, \theta)$: each new element joins an existing group with probability proportional to the group size (discounted by $\alpha$), or starts a new group with probability proportional to $\theta + \len \alpha$, where $\len$ is the current number of groups. See \citet{pitman2006combinatorial} for background.


\subsection{Coagulation algebra on partitions}
\label{subsec:coag}

To work with unlabelled partition-valued data, we introduce combinatorial operations that merge two partitions.
These operations constitute the algebraic mechanism underlying the finite mixture representations on which our inferential framework is built.

\begin{definition}[Indexed coagulation]\label{def:indexed-coag}
Let \(\omega \in \mathcal{P}_m\), \(\gamma \in \mathcal{P}_n\). An \emph{indexed coagulation} of $\omega$ and $\gamma$ assigns parts of \(\gamma\) to subsets of indices in \(\omega\), sums assigned entries, and retains all unassigned parts.
\end{definition}

Any indexed coagulation of two partitions thus produces a partition \(\mu\) coarsening their joint information. Fig.~\ref{fig:diagrams} (right) shows an instance with \(\omega = (2,1,1)\), \(\gamma = (3,2,1,1)\), yielding $\mu=(4,3,2,1,1)$. 
Informally, an indexed coagulation can be thought of as assigning ``colors'' to the parts of $\omega$ and $\gamma$ in a coordinated way, and then \emph{matching} (that is, merging) all parts that share the same color, while leaving uncolored parts unchanged.
See SM, Def.\ref{def:coag-supp}, for a more formal definition. 

The following generalizes Def.~\ref{def:indexed-coag}.

\begin{definition}[General coagulation]\label{def:coag}
Let \(\omega, \gamma \in \mathcal{P}\). The \emph{coagulation set} \(\coag(\omega, \gamma)\) collects all partitions \(\mu\) obtainable as indexed coagulations of \(\omega\) and \(\gamma\). For sets \(\Omega, \Gamma \subset \mathcal{P}\),
\[
\coag(\Omega, \Gamma) := \bigcup_{\omega \in \Omega,\, \gamma \in \Gamma} \coag(\omega, \gamma).
\]
\end{definition}

\begin{figure}[t!]
    \centering
    \includegraphics[width=\textwidth]{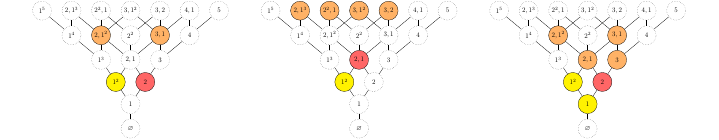}
\caption{\scriptsize
Examples of coagulation sets $\coag(\Omega,\Gamma)$ (orange) from $\Omega$ (yellow) and $\Gamma$ (red).
The notation $a^{b}$ means $b$ parts equal to $a$.
\emph{Left:} $\Omega=\{(1^{2})\}$, $\Gamma=\{(2)\}$.
\emph{Center:} $\Omega=\{(1,1)\}$, $\Gamma=\{(2,1)\}$.
\emph{Right:} $\Omega=\{(1),(1^{2})\}$, $\Gamma=\{(2)\}$; coagulation occurs only between $\Omega$ and $\Gamma$, hence $\mu = (3)$ arises from merging $\omega = (1)$ with $\gamma = (2)$.}
    \label{fig:coagulation}
\end{figure}

Informally, the coagulation set $\coag(\omega,\gamma)$ consists of all distinct partitions that can be formed by merging the blocks of $\omega$ and $\gamma$ in every possible way: one considers all possible matchings between blocks of the two partitions, merges matched blocks by summing their sizes, keeps unmatched blocks unchanged, and re-rank all the parts thus obtained in decreasing order.
Each $\mu \in \coag(\omega, \gamma)$ corresponds to one possible merging configuration obtained by matching parts of $\omega$ and $\gamma$ (Fig.~\ref{fig:coagulation}). 

We now illustrate how coagulation enters the predictive structure of the Ewens--Pitman model.
Specifically, we consider the conditional distribution of future clusters given an existing partition. Consider a CRP, and let $\Pi_{1:n}$ be the state of the process after the first $n$ customers have joined the restaurant, and $\Pi_{n+1:n+m}$ the marginal clustering of the next $m$ customers.
The following will recur multiple times in the coming sections.

\begin{proposition}[Conditional predictive distribution] 
\label{Conditional predictive distribution}
Let $\omega \in \mathcal{P}_n$, $\gamma \in \mathcal{P}_m$. Then
\begin{equation} \label{conditional CRP}
\psf^{\omega}(\gamma) := \Pr(\Pi_{n+1:n+m} = \gamma   |   \Pi_{1:n} = \omega)
= \sum_{\mu \in \coag(\omega, \gamma)} \H (\omega, \gamma   |   \mu) \frac{ \psf(\mu) }{\psf(\omega)}.
\end{equation}
When $\omega=\emptyset$, $\psf^{\omega} = \psf$.
The coefficients $\H(\omega, \gamma  |  \mu)$ are given in Proposition~\ref{prop:prod-cond-SM}.
\end{proposition}

Proposition~\ref{Conditional predictive distribution} shows that the conditional law of future clusters given an existing partition is a \emph{finite mixture} of Ewens--Pitman sampling formulas.
Each mixture component corresponds to a possible coagulation $\mu$ of the past partition $\omega$ with the future partition $\gamma$, and the weights $\mathcal H(\omega,\gamma | \mu)$ quantify the number of distinct matchings leading to the same merged configuration. In other words, predicting new clusters under the Ewens--Pitman model amounts to summing over all ways in which future observations can be attached to existing groups or form new ones, with probabilities determined by simple combinatorial counts. See the Supplement for a proof.


\subsection{Likelihood of partition-valued data}
\label{subsec:likelihood}

The Ewens--Pitman sampling formula in \eqref{PSF} describes the \emph{unconditional} law of an exchangeable random partition.
We now move to a \emph{conditional} setting, in which partitions are generated by sampling from a latent random probability vector $X(t)$ that evolves over time according to a diffusion.
This conditional formulation induces dependence between partitions observed at different times and leads to a hidden Markov model with infinite-dimensional latent state.

Let $\pi^{0:N} = (\pi^0, \ldots, \pi^N)$ denote partitions observed at times $t_0 < \cdots < t_N$.
The corresponding joint likelihood can be written as
\begin{equation}\label{eq:full-likelihood}
\Pr(\pi^{0:N}) = \int \cdots \int 
\left[ \prod_{k=1}^N P(\pi^k  |  x_{t_k})\, p(x_{t_k}  |  x_{t_{k-1}}) \right]
P(\pi^0  |  x_{t_0})p(x_{t_0})\, dx_{t_{0:N}},
\end{equation}
where $p(x_k  |  x_{k-1})$ is the transition density of the latent \emph{Poisson--Dirichlet diffusion}
(cf.~Sections~\ref{sec:model} and \ref{SM:petrov}),
and $P(\pi  |  x)$ is the likelihood of observing an unlabelled partition given $x$.

For $x = (x_{1}, x_{2}, \dots)$ in the decreasing infinite simplex and $\pi=(\pi_1,\ldots,\pi_{\len})$ a partition of $n$, the conditional likelihood takes the form
\begin{equation}\label{likelihood}
P(\pi  |  x)
= C(\pi)\!
\sum_{i_1 \ne \cdots \ne i_\len}
x_{i_1}^{\pi_1}\cdots x_{i_\len}^{\pi_\len},
\end{equation}
where $C(\pi)$ is the same combinatorial factor appearing in \eqref{PSF}.
The sum ranges over all injective assignments of latent labels to the blocks of $\pi$, reflecting the fact that the observation records only the \emph{group sizes} and not their labels.

Combined with the absence of a closed-form expression for the transition density
$p(x_{t_k}  |  x_{t_{k-1}})$
(see Eq.~\eqref{transition} in the Supplement),
the likelihood \eqref{eq:full-likelihood} is analytically intractable.
This motivates the search for algebraic identities that allow exact manipulation of the unlabelled likelihoods $P(\cdot  |  x)$.

Mirroring the coagulation-based identities obtained for the unconditional Ewens--Pitman model in Proposition~\ref{Conditional predictive distribution},
the following result provides the key product expansion for the conditional likelihood.

\begin{proposition}[Product expansion]\label{prop:prod-cond-main}
For any $x$ in the decreasing infinite simplex,
\begin{equation}
P(\omega  |  x)\, P(\gamma  |  x)
= \sum_{\mu \in \coag(\omega,\gamma)}
\H(\omega,\gamma  |  \mu)\, P(\mu  |  x),
\end{equation}
with explicit coefficients $\H(\omega,\gamma  |  \mu)$
(see Proposition~\ref{prop:prod-cond-SM}).
\end{proposition}

Thus, although the likelihood $P(\cdot  |  x)$ is not closed under multiplication,
products of unlabelled likelihoods admit a finite mixture representation over coagulated partitions.
This property yields a form of \emph{quasi-conjugacy}.

Together with Proposition~\ref{Conditional predictive distribution},
this expansion leads to exact, closed-form predictive distributions under the Ewens--Pitman model,
expressed as finite mixtures indexed by coagulated partitions.
The same mixture structure underlies the filtering and smoothing recursions developed in Section~\ref{sec:filtering}.
Their temporal propagation is governed by a dual pure-death process \citep{griffiths2024dual},
which we exploit to obtain recursive exact inference;
additional details are provided in the Supplement.


\section{Filtering two-parameter Poisson--Dirichlet HMMs} \label{sec:filtering}

\subsection{Model structure and latent process dynamics}
\label{sec:model}

We now formalize the statistical model.
The latent process is a two-parameter Poisson--Dirichlet diffusion
$X := (X(t), t \ge 0)$ with parameters $(\alpha,\theta)$, where
$0 \le \alpha < 1$ and $\theta \ge -\alpha$.
This infinite-dimensional process models time-evolving frequencies of
unobserved categories, such as species proportions or cluster sizes in graphs
(cf.~Fig.~\ref{fig:infectious}), with continuous paths in the ordered infinite
simplex, that is, decreasingly ordered vectors summing to one.
We assume $X(0) \sim \PD$, the two-parameter Poisson--Dirichlet \emph{distribution}, which is the stationary law of the diffusion.
It is defined as the law of the decreasing rearrangement $(W_{1}, W_{2}, \dots)^{\downarrow}$ of the stick-breaking process:
\begin{equation} \label{stick-breaking}
	W_{1} := V_{1}, \quad W_{i} := V_{i} \prod_{j < i} (1 - V_{j}) \quad \text{for} \ i \ge 2, \quad V_{i} \overset{ind}{\sim} \text{Beta}(1 - \alpha, \theta + i \alpha).
\end{equation}
See Section~\ref{SM:petrov} for background on the PD diffusion.

At discrete times $t_0 < t_1 < \dots < t_N$, we observe unlabelled partitions
$\Pi^k \in \mathcal P_{n_k}$ obtained by drawing $n_k$ samples from the latent
state $X(t_k)$ and grouping them by type, so that only block sizes are recorded
and labels are ignored (Fig.~\ref{fig:diagrams}, left).
The observation (emission) model is
\begin{equation}\label{eq:hmm-structure}
\Pi^k  |  X(t_k) \sim \paint(n_k, X(t_k)),
\end{equation}
where $\paint(n,x)$ denotes Kingman's paintbox construction.
Equation~\eqref{eq:hmm-structure} specifies only the conditional law of the
observations given the latent state.

While the Poisson--Dirichlet diffusion is stationary and hence
$X(t_k)\sim\PD$ marginally for each $k$, temporal dependence across time is
induced by the diffusion dynamics described in
\eqref{eq:latent-transition-hierarchy}.
In particular, the conditional law of $X(t_k) |  X(t_{k-1})$ is not
Poisson--Dirichlet.

Conditionally on the latent trajectory $X(t_0),\dots,X(t_k)$, the observations
$\Pi^{0:k}$ are independent, with distribution given by the emission model \eqref{eq:hmm-structure}.
Under the paintbox construction, samples are independently assigned to indices
$i\in\mathbb N$ with probabilities $x_i$ and grouped by common index; the
resulting unlabelled partition records only block sizes.
All permutations of labels yielding the same block structure are therefore
equivalent.
Accordingly, the probability that a random partition
$\Pi\sim\paint(n,x)$ equals a given $\pi\in\mathcal P_n$ is the symmetric
function $P(\pi |  x)$ in \eqref{likelihood}.

Marginalizing over $X(t_k)$ in \eqref{eq:hmm-structure} yields $\Pi^k \sim \psf$,
a classical consequence of Kingman's representation theorem
\citep{kingman_random_1975,kingman_representation_1978};
see Section~\ref{SMsec:partition-structures}.

This defines a nonparametric hidden Markov model with infinite-dimensional
latent states and discrete, combinatorial observations.
We next describe the latent signal dynamics and explain why direct inference
based on these transitions is analytically intractable.

\paragraph{Latent dynamics.}
The hidden Markov structure of the model is completed by specifying the
\emph{transition mechanism} of the latent process.
Although the Poisson--Dirichlet diffusion is stationary, so that
$X(t_k)\sim\PD$ marginally for each $k$, the conditional law of
$X(t_k) |  X(t_{k-1})$ is not Poisson--Dirichlet.
Instead, the transition kernel of the diffusion admits the following
hierarchical representation in terms of an auxiliary latent partition:
\begin{equation} \label{eq:latent-transition-hierarchy}
\begin{aligned}
&N_k \sim \mathcal{B}_\theta(\Delta_k), \qquad \Delta_k := t_k - t_{k-1},\\
&(\lambda(t_{k})    |   N_k = n,\; X(t_{k-1}) = x) \sim \paint(n, x), \\
&(X(t_k)   |   \lambda(t_{k}) = \lambda) \sim \PD^\lambda.
\end{aligned}
\end{equation}
Interpretation is as follows.
An integer $N_k$ is drawn from $\mathcal{B}_\theta(\Delta_k)$, the distribution
of the number of blocks in Kingman's coalescent, a pure-death process starting
from infinity (Section~\ref{SM:petrov}).
Conditional on $N_k=n$, $n$ latent samples are drawn from $X(t_{k-1})$ to induce
a latent partition $\lambda(t_k)\sim\paint(n,x)$.
Given this latent partition, the state $X(t_k)$ is distributed according to the
posterior law $\PD^\lambda$.
Thus, the Markov dependence of $X(t_k)$ on $X(t_{k-1})$ is mediated by the latent
partition $\lambda(t_k)$, even though the marginal distribution of $X(t_k)$
remains $\PD$.

Note that $N_k$ depends on $\Delta_k$ but not on $x$, and governs temporal dependence: small $\Delta_k$ yields large $N_k$ and strong dependence, with $\Delta_k \to 0$ recovering continuity, while large $\Delta_k$ induces near-independence, $\Delta_k \to \infty$ leading to stationarity ($X(t_k) \sim \PD$). See Section~\ref{SM:petrov} for details.

The posterior law $\PD^\lambda$ was characterized by \citet{pitman96} in the \emph{labelled} case; see \citet[][Prop.~4.1]{griffiths2024dual} for the \emph{unlabelled} case.
If $X \sim \PD$ and $\Pi   |   X \sim \paint(n, X)$, then for any particular $\pi = (\pi_{1}, \dots, \pi_{\len})  \in \P_{n}$
\begin{equation}\label{posterior tppd}
(X   |   \Pi = \pi) \overset{d}{=} ((1-W) V^{(1)}_1, \dots, (1-W) V^{(1)}_{\len}, W V^{(2)}_1, W V^{(2)}_2, \dots)^\downarrow,
\end{equation}
where $(\cdot)^{\downarrow}$ denotes decreasing rearrangement, and the components are independent:
\begin{equation}\nonumber
W \sim \mathrm{Beta}(\theta + \len \alpha, n - \len \alpha ), \quad
V^{(1)} \sim \mathrm{Dirichlet}(\pi_1 - \alpha, \dots, \pi_{\len} - \alpha), \quad
V^{(2)} \sim \text{PD}_{\alpha, \theta + \len \alpha }.
\end{equation}
Although not a $\PD$, we denote the law of $X   |   \Pi = \pi$ in \eqref{posterior tppd} by $\PD^\pi$ for notational convenience. 
Sampling from $\PD^\pi$ is feasible via Algorithm~\ref{alg: conditional PD}, which relies on $\varepsilon$-truncated simulation methods (Algorithm~\ref{alg: truncated PD}). See Section~\ref{SM:alg}.

\begin{remark}\label{singleton-exception}
A singleton makes an exception, since $\PD^{(1)}\overset{d}{=}\PD$; this is peculiar to unlabelled models, since an observed singleton carries no information on the grouping structure.
\end{remark}

In the next section we show that, when observations are incorporated, the joint
law of the latent state $X(t_k)$ and the data $\Pi^{0:k}$ admits a conditional
independence representation given a latent coagulation.
This yields a posterior inference scheme involving only finitely many
computations, in contrast to the infinite-dimensional transition dynamics in
\eqref{eq:latent-transition-hierarchy}, which describe the evolution of the
latent state independently of the data.

\paragraph{Notation for conditional laws.}
We let $\nu_{k|h:j}(A) := \Pr(X(t_k) \in A   |   \Pi^h, \dots, \Pi^j)$ denote the law of $X(t_k)$ given observations in $[t_h, t_j]$. This accommodates:
filtering: $k = j$ (online estimation);
 prediction: $k > j$ (forecasting);
 smoothing: $k < j$ (offline estimation).


\subsection{Posterior inference: filtering, smoothing, and forecasting} \label{sec:posterior-inference}

Direct use of the transition dynamics \eqref{eq:latent-transition-hierarchy} is intractable. We instead exploit a \emph{dual} representation of the \tppd diffusion in terms of a pure-death process $N(t)$ with finite starting point \citep{griffiths2024dual}. This duality allows the transition $X(t_k) |  X(t_{k-1})$ to be expressed, conditionally on observed data, through a latent partition with finite support, yielding tractable recursive filtering.

We now summarize the main inferential results for the \tppd HMM \eqref{eq:hmm-structure}, with full details in the Supplement (Sections~\ref{SMsec:prediction}-\ref{SMsec:smoothing}).

\paragraph{General posterior representation.}
All posterior laws of interest, including filtered, smoothed, and interpolated distributions, belong to the following finite mixture family:
\begin{equation} \label{finite-mixture-family}
\mathcal{F} := \left\{ \nu = \sum_{\lambda \in \Lambda} w_\lambda \PD^\lambda :\  \Lambda \subset \mathcal{P},\ |\Lambda| < \infty,\ \sum_{\lambda \in \Lambda} w_\lambda = 1 \right\},
\end{equation}
with each $\PD^\lambda$ as in \eqref{posterior tppd} and weights $w_\lambda$ being explicit functions of the observations.

The key structural feature is that $\mathcal{F}$ is closed under prediction and update, yielding recursive filtering and smoothing procedures. Thus, despite the infinite-dimensional nature of the signal, all posterior distributions for $X(t)$ evolve within a finite-dimensional, tractable space. 
Model \eqref{eq:hmm-structure} is thus \emph{quasi-conjugate}, in the sense that at any $t$, unconditionally we have $X(t) \sim \PD$, while given observed data
\begin{equation} \label{eq:posterior-general}
X(t)   |   \Pi^{0:N} \sim \sum_{\lambda \in \Lambda_t} w_\lambda(t)\, \PD^\lambda,
\end{equation}
with $\Lambda_t \subset \mathcal{P}$ being a finite and deterministic set determined by the data. This quasi-conjugacy mirrors the role of conjugate priors in parametric models: although the likelihood is not closed under multiplication, Proposition~\ref{prop:prod-cond-main} shows that products of likelihoods decompose as finite mixtures of the same kernel, preserving tractability.

Equivalently, \eqref{eq:posterior-general} corresponds to the hierarchy
\[
X(t)   |   \lambda(t) = \lambda \sim \PD^\lambda, \qquad \lambda(t)   |   \Pi^{0:N} \sim w_\bullet(t),
\]
where $w_\lambda$ has finite support and reflects the full observation history. The evolution of these latent quantities (sets $\Lambda_{t}$ and weights $w_\lambda(t)$) is governed explicitly by the dual process and the coagulation algebra introduced in Section~\ref{subsec:coag}.

\paragraph{Online estimation (filtering).}
We compute the filtered distribution of the latent state at observation time $t_k$ as a finite mixture
\begin{equation} \label{eq:filter-mixture}
\nu_{k|0:k} = \Pr(X(t_k) \in \cdot   |   \Pi^{0:k}) = \sum_{\lambda \in \Lambda_{0:k}} w_\lambda\, \PD^\lambda,
\end{equation}
where $\Lambda_{0:k} \subset \mathcal{P}$ is a finite collection of partitions determined by the data collected up to time $t_{k}$, and $w_\lambda$ are explicitly computable data-dependent weights (we will drop the dependence of $w_{\lambda}(t)$ on $t$  for notational convenience when this causes no confusion). The posterior retains the structure of a finite mixture of conditionally conjugate components $\PD^\lambda$, which is preserved under recursive prediction and update operations. Define the \emph{lower set} of $\Lambda \subset \P$ to be
\begin{equation}\label{lower set}
L(\Lambda) = \{\omega : \omega \sp \lambda, \lambda \in \Lambda \}, 
\end{equation} 
i.e., the set of partitions that can be obtained by removing units from partitions in $\Lambda$.

\begin{proposition}[Recursive filtering structure] \label{prop:filtering}
Let the latent state $X(t_{k-1})$ conditional on the partitions $\Pi^{0:k-1}$ observed up to time $t_{k-1}$ have posterior distribution $\nu_{k-1|0:k-1}= \sum_{\lambda \in \Lambda } w_\lambda \PD^\lambda$. Then, at time  $t_k$:
\begin{enumerate}[(i)]

\item \textbf{Prediction}. The law $\nu_{k|0:k-1}$ of the latent state $X(t_k)$ conditional on the partitions $\Pi^{0:k-1}$ observed up to time $t_{k-1}$ belongs to $\F$. In particular, 
\begin{equation}\label{propagation described}
    X(t_k)  |  \lambda(t_{k}) \sim \PD^{\lambda(t_{k})}, \qquad
     \lambda(t_{k}) \sim \sum_{\lambda\in \Lambda}w_{\lambda}\dualtr_{\lambda, \bullet}(\Delta_{k}),
\end{equation}
where $\dualtr_{\lambda, \bullet}(\Delta_{k})$ are the transition probabilities of the dual (cf.~\eqref{dual transition}). 

\item \textbf{Update}. Let $v_\omega := \Pr(\lambda(t_{k}) = \omega)$ denote the distribution of $\lambda(t_k)$ in \eqref{propagation described}, and let $\Pi^k = \pi^k$ be the observed partition at time $t_{k}$. Then the law $\nu_{k|0:k}$ of the latent state $X(t_k)$ conditional on the partitions $\Pi^{0:k}$ observed up to time $t_{k}$ belongs to $\F$. In particular, 
\begin{equation}\label{updated distribution described}
    X(t_k)  |  \lambda(t_{k}) \sim \PD^{\lambda(t_{k})}, \qquad \lambda(t_{k})  |  \pi^k \sim \sum_{\omega \in L(\Lambda)}  \hat v_\omega \Coag(\omega, \pi^k),
\end{equation}
where $\hat v_\omega \propto v_\omega \psf^{\omega}(\pi^{k})$, $\psf^{\omega}$ is as in 
\eqref{conditional CRP}, and $\Coag(\omega, \pi^k)$ is fully described in \eqref{mu-given-omega-gamma}.
\end{enumerate}
\end{proposition}

This result highlights the key structural property of the filtering scheme: the latent state $X(t_k)$, conditioned on data $\Pi^{0:k}$, always admits a finite mixture representation over posterior distributions $\PD^\lambda$. See Theorems \ref{thm:propagation} and \ref{thm:update} for alternative formulations equivalent to \eqref{propagation described} and \eqref{updated distribution described}.

The recursive construction unfolds over two alternating stages, each preserving membership in the mixture family $\F$:
\begin{itemize}
\item The \emph{prediction step} (part (i)) advances the prior distribution from $t_{k-1}$ to $t_k$ via the dual block-counting process. This process propagates the latent partition $\lambda(t_{k-1})$ forward in time, producing a new latent partition $\lambda(t_{k})$ according to the death process transitions $\dualtr_{\lambda,\bullet}(\Delta_k)$. Conditionally on $\lambda(t_{k})$, the state $X(t_k)$ is distributed as $\PD^{\lambda(t_{k})}$, preserving analytic tractability.
\item The \emph{update step} (part (ii)) incorporates the new observation $\Pi^k = \pi^k$ by modifying the law of the latent partition $\lambda(t_{k})$ through the coagulation operation. Each predicted partition is merged with the observed data to produce an updated pool of information.
Again, conditionally on $\lambda(t_{k})$, the law of $X(t_k)$ remains in the conjugate class $\PD^{\lambda(t_{k})}$.
\end{itemize}

In both steps, the latent partition $\lambda(t_{k})$ plays a central role: it acts as a sufficient statistic for the posterior component, while its evolution is governed by the dual process (for prediction) and by the algebra of coagulations (for updates). Crucially, the distribution of the latent partition  summarizes the relevant information accrued up to time $t_k$. Hence, filtering in the \tppd model can be interpreted as tracking the evolution of this latent partition, rather than the infinite-dimensional law of $X(t)$ directly.

Moreover, the recursion is closed in the sense that it only involves a finite and tractable number of latent partitions at each time step. The support $\Lambda_t$ of the mixture evolves deterministically, and the updated weights can be computed explicitly. This renders the inference scheme fully analytic and computationally efficient, with no need for Monte Carlo methods.

In summary, the recursive filtering algorithm maintains tractability by leveraging the dual structure of the PD diffusion and the partition algebra. The hidden state $X(t_k)$ is always conditionally distributed as a finite mixture of $\PD^\lambda$ distributions, where each $\lambda$ reflects a possible state of a latent, unobserved partition $\lambda(t_{k})$ guided by past data and current observations.

\begin{figure}[t!]
    \centering
    \includegraphics[width=\textwidth]{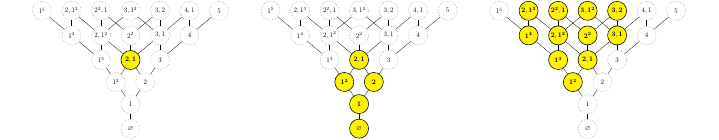}
    \caption{\scriptsize
Latent partition support in recursive filtering via prediction and update. {Left:} initial support $\Lambda = \{(2,1)\}$ at $t_0$. {Center:} prediction expands $\Lambda$ this to $L(\Lambda)$ using the dual process. {Right:} update with $\pi^1 = (1,1)$ produces $\coag(L(\Lambda), \pi^1)$ as the new support. Each node represents a component $\PD^\lambda$.}
    \label{fig:lower set}
\end{figure}

Fig.~\ref{fig:lower set} illustrates this process: an initial observed partition $\pi^0 = (2,1)$ (left graph) yields posterior $X(t_0) \sim \PD^{(2,1)}$. The prediction step computes a lower set $L((2,1))$ via the dual death process (center graph), forming the support of the latent partitions in $\nu_{1|0}$. Then, observing $\pi^1 = (1,1)$ leads to the updated support $\coag(L((2,1)), (1,1))$ (right graph) through the product expansion. The recursion continues analogously at later times.

This structure enables tractable online inference: filtering requires only propagation and update of a finite set of partitions $\lambda$ and their weights $w_\lambda$, with no need for simulation or numerical integration.

\paragraph{Offline estimation (smoothing).}
While filtering tracks the evolving law of the latent state $X(t_k)$ using data up to time $t_k$, smoothing provides retrospective estimates by conditioning on the full observation sequence $\Pi^{0:N}$. The goal is to compute the distribution of $X(t_k)$ given all data, i.e.,
$\nu_{k|0:N} = \Pr(X(t_k) \in \cdot   |   \Pi^{0:N}),$
for arbitrary $0 \le k \le N$.

The key insight is that smoothing retains the same latent partition structure as filtering: at any time $t_k$, the smoothed distribution of $X(t_k)$ remains a finite mixture over conditionally conjugate components $\PD^\mu$, where each index $\mu$ reflects a plausible state of the latent partition consistent with both past and future data. As in filtering, the latent partition acts as sufficient statistics for posterior inference, and their evolution is governed by algebraic operations on partitions.

Smoothing is achieved by combining:
(i) the forward information from the filter at time $t_k$, encoded by a finite collection $\Lambda_{0:k}$ of partitions and their associated weights; and (ii) future information propagated backwards from time $t_{N}$ up to $t_{k}$, encoded by a finite support $\Omega^{k:N}$ of partitions that are compatible with future data.

At a high level, the smoothing algorithm proceeds as follows:
\begin{list}{$\bullet$}{
\itemsep=0mm \topsep=2mm \itemindent=-3mm \labelsep=2mm \labelwidth=0mm
\leftmargin=9mm \listparindent=0mm \parsep=0mm \parskip=0mm \partopsep=0mm
}
\item Run the forward filtering recursion up to time $t_k$ to obtain $\Lambda_{0:k}$
\item Run a backward pass from $t_N$ down to $t_{k+1}$, and propagate further to $t_{k}$, constructing a compatible set of backward partitions $\Omega^{k:N}$.
\item For each pair $(\lambda, \omega) \in \Lambda_{0:k} \times \Omega^{k:N}$, form a new set of partitions $\coag(\lambda, \omega)$.
\item Assign to each coagulation $\mu$ thus obtained, hence to $\PD^\mu$, a weight computed from the product of the forward and backward contributions.
\end{list}

The resulting posterior distribution is a finite mixture over partitions $\mu$, each representing a coherent reconstruction of the system state at time $t_k$ consistent with the full dataset.

\begin{figure}[t!]
    \centering
    \includegraphics[width=\textwidth]{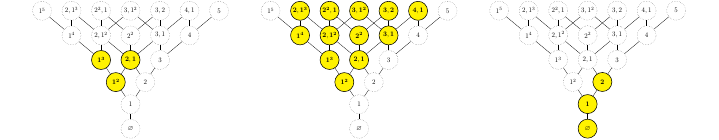}
    \caption{\scriptsize
    Smoothing via forward-backward latent structure. The \emph{forward} latent partition support $\Lambda_{0:k}$ (left) is coagulated with the \emph{backward} latent partition support $\Omega^{k:N}$ (right), to form the coagulation set $\coag(\Lambda_{0:k}, \Omega^{k:N})$ (middle). The latter supports the latent partitions in the posterior law of $X(t_k)  |  \Pi^{0:N}$.}
    \label{fig:smoothing lattice}
\end{figure}

Despite the bidirectional nature of the algorithm, tractability is fully preserved. All steps involve only finite sets of partitions, and weights can be computed exactly using closed-form expressions. No simulation or numerical integration is required. Full mathematical details, including the mixture representation and formal proof of correctness, are provided in Theorem~\ref{thm:smoothing} of the Supplement.

\paragraph{Forecasting and interpolation.}
The same recursive structure extends naturally to inference at unobserved time points. For forecasting beyond the last observation time $t_N$, the distribution of the latent state $X(t)$ for $t > t_N$ is obtained by propagating the filtered distribution $\nu_{N|0:N}$. Thanks to Theorem~\ref{thm:propagation} in the SM, this again takes the form \eqref{eq:posterior-general}, where the weights $w_\lambda(t)$ reflect the evolution of the dual process beyond the observation window. E.g., if at $t_{N}$ the latent partition has support as in the rightmost graph of Fig.~\ref{fig:lower set}, call it $\Lambda_{0:N}$, then $\Lambda_{t}$ in \eqref{eq:posterior-general} is the lower set $L(\Lambda_{0:N})$, which in this case is the union of $\Lambda_{0:N}$ with $\{(3), (2), (1), \emptyset\}$.
 As the forecast horizon grows ($t \to \infty$), the probability mass over $L(\Lambda_{0:N})$ progressively concentrates towards nodes in a neighborhood of $\emptyset$, implying that the influence of past data vanishes and the forecast distribution converges to the stationary law $\PD$ (cf.~Corollary~\ref{cor: convergence to stationarity})

Finally, if one seeks to interpolate between observation times to estimate $X(t)$ for $t \in (t_k, t_{k+1})$, the posterior takes again a qualitatively similar form, but the weights depend explicitly on the time offset $t - t_k$. This is formalized in Corollary~\ref{cor:interpolation}, enabling smoothed inference at arbitrary intermediate time points.

\paragraph{Predictive functionals.} 
A byproduct of the above results is that the law of future or interpolated data $\Pi'(t)$ induced by $m$ samples from $X(t)$ is a finite mixture of predictive CRP distributions. Specifically
\begin{equation} \label{predictive}
\Pi'(t)  |  \lambda(t) = \lambda \sim \psf^{\lambda}, \quad \lambda(t)  |  \Pi^{0:N} \sim w_{\bullet}(t)
\end{equation}
with $\psf^{\lambda}$ as in \eqref{conditional CRP}.
In particular, if $t > t_{N}$, this is achieved by evolving the current sufficient statistic \(\lambda(t_{N})\) forward in time through the dual process to obtain $\{w_\lambda(t), \lambda \in \Lambda_t\}$. Then, conditional on $\lambda(t) = \lambda$, a future observation $\Pi'(t)$ has law $\psf^{\lambda}$.
This yields interpretable predictions and simulations of future observations.

The finite-mixture representation \eqref{eq:posterior-general} also supports principled forecasting and uncertainty quantification for functionals of the latent random measure $X(t)$, such as diversity indices \citep{patil1982diversity}.
For example, heterozygosity has a natural interpretation as the probability that two samples drawn from $X(t)$ belong to different types. 
Under the finite-mixture representation, the posterior distribution of heterozygosity at time $t$ is itself a finite mixture of the corresponding distributions induced by the $\PD^\lambda$ laws, from which sampling is straightforward.
This strategy is implemented in Section~\ref{sec:num_exp}.

We summarize the above results in the following theorem.

\begin{theorem}[Exact inference for PD-HMMs]
\label{thm:main-inference}
Let $X = (X(t), t \ge 0)$ be a $\PD$-diffusion observed via partition-valued data $\Pi^0, \dots, \Pi^N$ as in \eqref{eq:hmm-structure}, and let $\F$ be as in \eqref{finite-mixture-family}. Then:
\begin{list}{
\arabic{enumi}.
}{\itemsep=1mm\topsep=2mm\itemindent=-3mm\labelsep=2mm\labelwidth=0mm\leftmargin=5mm\listparindent=0mm\parsep=0mm\parskip=0mm\partopsep=0mm\rightmargin=0mm\usecounter{enumi}}
\setcounter{enumi}{0}
\item  \textbf{Filtering and prediction.} For each $k \le N$ and $h \ge 0$, the filtering and forecast distributions $\nu_{k|0:k}$ and $\nu_{k+h|0:k}$ belong to the family $\mathcal{F}$, and are computable via a recursive procedure based on finite-lattice transitions (cf.~Theorems~\ref{thm:propagation}-\ref{thm:update}).

    \item \textbf{Smoothing.} For each $k < N$, the smoothed distribution $\nu_{k|0:N}$ also lies in $\mathcal{F}$ and is obtained through a backward-forward recursion over partitions (cf.~Theorem~\ref{thm:smoothing}).

    \item \textbf{Forecasting new data.} For any $t \in [0, \infty)$ and $n \in \mathbb{N}$, the conditional distribution of a new data point $\Pi'(t)$ given $\Pi^{0:N}$ is a finite mixture of conditional Ewens--Pitman sampling formulae (cf.~Corollary~\ref{cor: forecasting}).
\end{list}
\end{theorem}

\begin{figure}[t!]
\centering
\includegraphics[width=.9\linewidth]{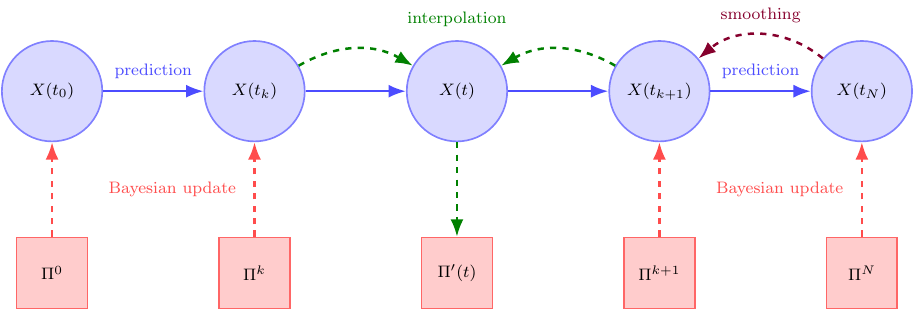}
\caption{\footnotesize Recursive inference in the hidden PD model. Arrows denote prediction (solid blue), Bayesian update (dashed red), interpolation (dashed green), and backward smoothing (dashed purple).}
\label{fig:recursive-inference}
\end{figure}

Theorem~\ref{thm:main-inference} establishes that posterior inference for the hidden Markov model \eqref{eq:hmm-structure} is tractable: all relevant conditional distributions admit closed-form finite-mixture representations of quasi-conjugate components. The associated mixture weights and simulation algorithms are detailed in the Supplement (cf.~Algorithms~\ref{alg: signal prediction}--\ref{alg: forecasting}). 
This yields exact recursive inference for infinite-dimensional nonparametric latent dynamics without resorting to MCMC or SMC methods. 
Fig.~\ref{fig:recursive-inference} illustrates the resulting framework, which unifies prediction, filtering, smoothing, and interpolation.

\section{Model Implementation}
\label{sec: illustration}

\subsection{Summary of the algorithms}
\label{sec:algorithm}

The full filtering and smoothing procedures, together with all intermediate
computational steps, are detailed in the pseudocode provided in
Section~\ref{SM:alg}.  
In particular, Algorithms~\ref{alg: truncated PD} and
\ref{alg: conditional PD} are auxiliary routines for simulating from an
$\varepsilon$-truncated $\PD$ distribution and from a conditional
$\varepsilon$-truncated $\PD^\pi$, respectively.  
Here, $\varepsilon$-truncation refers to terminating the stick-breaking
construction in \eqref{stick-breaking} once the remaining (residual) mass falls below $\varepsilon$,
thereby yielding a finite-dimensional approximation of the
Poisson--Dirichlet law.
The core inferential operations are handled by the remaining algorithms: Algorithm~\ref{alg: signal prediction} samples from the propagation step (cf.~\eqref{propagation described}); Algorithm~\ref{alg: updated prediction} covers the update step (cf.~\eqref{updated distribution described}); and Algorithm~\ref{alg: generic filtering} combines these components to perform online filtering via the recursive operations (cf.~\eqref{recursion}). For offline inference, Algorithms~\ref{alg: smoothing} and~\ref{alg: forecasting} support latent state estimation and partition simulation, respectively, after observing the full dataset.


\subsection{Computability and algorithmic strategies}

When initialized at \(X(0) \sim \PD\), the latent system satisfies \emph{computability} in the sense of \citet{chaleyat2006computable}: all conditional distributions of \(X(t)\) lie in the finite mixture family \(\mathcal{F}\) of Equation~\eqref{finite-mixture-family}, and thus admit exact evaluation without resorting to MCMC or SMC.

Nevertheless, efficient approximations are often desirable. Two key aspects are addressed in our implementation. First, computing the mixture weights \(\dualtr_{\lambda, \bullet}(t)\) in Equation~\eqref{propagation} (the transition probabilities of a pure-death process on $\mathcal{P}$ related to Kingman's coalescent in Eq. \eqref{eq:latent-transition-hierarchy}) can be computationally intensive. Although exact formulas are available, we use a Monte Carlo approach based on the Gillespie algorithm \citep{gillespie2007stochastic} to simulate paths of the death process (see Algorithm~\ref{alg: signal prediction gillespie}). This approximation may encounter numerical difficulties when starting from large partitions and propagating over short time intervals; in such cases, Gaussian approximations \citep{griffiths1984asymptotic} provide a useful alternative.

The second computational challenge arises from the rapid growth of the partition space as sample size increases. However, even in the unlabelled setting, most of the mixture's probability mass is typically concentrated on a small subset of components. Fig.~\ref{fig:mixture cardinality} illustrates this concentration, showing that mixtures pruned to include only components with cumulative weight above a chosen threshold (e.g., 95\%) retain most of the mass. Accordingly, we adopt pruning techniques as in \citet{king2021exact, king2024approximate}, either by retaining components up to a fixed cumulative mass or selecting the top components by weight.

\begin{figure}
    \centering
    \includegraphics[width=\linewidth]{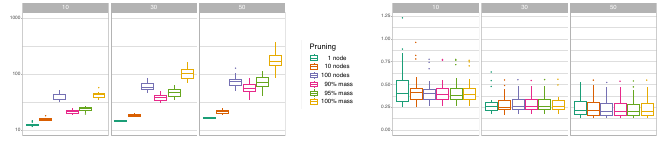}
    \caption{\scriptsize Left: Runtime (seconds, log scale) of Algorithm~\ref{alg: generic filtering} for sample sizes \(|\gamma| = 10, 30, 50\), and various pruning strategies (top-weighted components or total mass thresholds). Filtering draws \(10^4\) samples after each update using Algorithm~\ref{alg: signal prediction gillespie}. Parameters: \(\alpha = 0.1\), \(\theta = 1.5\), \(N = 9\), \(\Delta_k = 0.2\) for all \(k\), with \(\PD\) truncation threshold \(\varepsilon = 0.005\). Right: Associated negatively oriented interval scores (Equation~\eqref{eq: score}) based on 95\% credible intervals for heterozygosity.}
    \label{fig:time_vs_pruning_well_specified}
\end{figure}

As shown in Fig.~\ref{fig:time_vs_pruning_well_specified}, substantial speed gains result from pruning, especially at larger sample sizes. Pruning to 90\% total mass or retaining 10 components delivers near-identical accuracy compared to using the full mixture. The accuracy here is quantified by the \emph{negatively-oriented interval scores} \citep{gneiting2007strictly}
\begin{equation} \label{eq: score}
    \mathcal{S}^{(N)} = \frac{1}{N} \sum_{k = 0}^N \mathcal{S}_{\kappa}(l_{t_k}, u_{t_k}, y_{t_{k}}),
\end{equation}
where \(\mathcal{S}_\kappa(l, u, y) = (u - l) + \frac{2}{\kappa}(l - y) \mathbbm{1}\{y < l\} + \frac{2}{\kappa} (y - u) \mathbbm{1}\{y > u\}\), and \(l,u\) are the \(\kappa/2\) and \(1 - \kappa/2\) quantiles of the filtering distribution for a functional quantity \(y\). This score measures forecast accuracy by penalizing both the width of the predictive interval and any observations that fall outside it, with lower values indicating better performance. We apply this to the heterozygosity \(H_2(x) := 1 - \sum_{i \ge 1} x_i^2\), comparing its true value with the 95\% credible intervals from the filter. Our results support pruning as a valid and efficient approach for practical implementation.


\subsection{Parameter estimation}\label{sec:par estim}

Joint inference for $(\alpha,\theta)$ in the $\text{PD}_{\alpha,\theta}$ model is nontrivial, even in static or labelled settings (see \citealp{balocchi2026, cereda2023learning}). In our dynamic unlabelled setting, we estimate these parameters via maximum likelihood using the sequence of filtering distributions, thereby directly targeting the evolving latent trajectory.

The marginal likelihood \eqref{eq:full-likelihood} of the observed data sequence is given by
\begin{equation}
    \Pr(\Pi^0 = \pi^0, \dots, \Pi^N = \pi^N) = \psf(\pi^0) \prod_{k = 1}^{N} \sum_{\omega \in L(\Lambda_{0:k-1})} v_{\omega} \psf^{\omega}(\pi^{k})
\end{equation}
where the weights $v_{\omega}$ are those of the propagated law of the latent partition as given in Proposition \ref{prop:filtering}; cf.~\eqref{PSF} and \eqref{conditional CRP}. 
Each term on the right-hand side corresponds to a predictive partition distribution, which can be computed from Corollary~\ref{cor: forecasting} and is readily available from the weights computed during the update step in Algorithm~\ref{alg: updated prediction}. Thus, maximum likelihood can be implemented as a by-product of online inference.


\subsection{Numerical experiments}\label{sec:num_exp}

We illustrate our methodology using synthetic data from a known $\PD$ diffusion trajectory. The goal is to estimate the heterozygosity functional of the latent state $x$, given by $H_2(x) := 1 - \sum_{i \ge 1} x_i^2$. Although we focus on $H_2$, our approach can be readily adapted to other functionals on the ordered infinite simplex, such as entropy and diversity indices \citep{patil1982diversity}.

A latent trajectory is simulated at $20$ equispaced time points ($\Delta_k = 0.025$) with $(\alpha, \theta) = (0.1, 1.5)$, and observations are generated as unlabelled partitions of size 50 from model~\eqref{eq:hmm-structure}. Filtering and smoothing are performed via Algorithms~\ref{alg: generic filtering} and \ref{alg: smoothing}, using pruning to retain the 10 mixture components with largest weights after each propagation and update step. Gillespie-based propagation uses $10^4$ particles (Algorithm~\ref{alg: signal prediction gillespie}), and posterior summaries are computed from $10^4$ posterior draws.

\begin{figure}[t!]
    \centering
    \includegraphics[width=.9\linewidth]{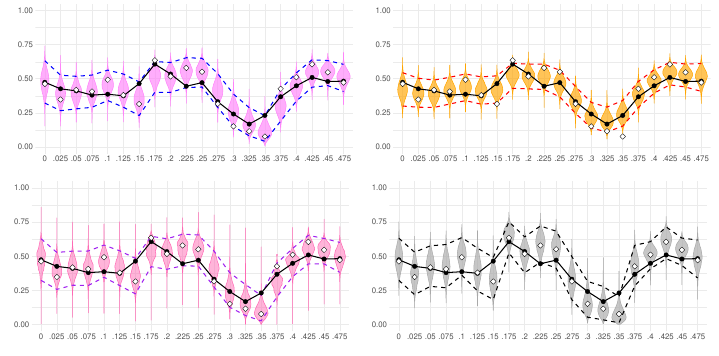}
    \caption{\scriptsize Posterior inference for heterozygosity using filtering (top left), smoothing (top right), bootstrap particle filter (bottom left) and independent priors (bottom right). Black line: true $H_2$ trajectory; white diamonds: observed heterozygosity $\hat H_2(\pi^k)$. Violin plots show posterior densities; dashed lines show 95\% credible intervals.}
    \label{fig:entropy2_filter_smooth_mle}
\end{figure}

Model parameters are estimated by maximizing the marginal likelihood over a discrete grid, yielding $(\hat{\alpha}, \hat{\theta}) = (0, 1.5)$, underestimating $\alpha$. Fig.~\ref{fig:entropy2_filter_smooth_mle} displays marginal posterior estimates for $H_2(x)$ at the observed time points. Violin plots represent posterior densities (via kernel density estimation), triangles mark 95\% credible intervals, and white diamonds indicate the observed heterozygosities $\hat H_2(\pi^k) = 1 - \sum_i (\pi_i^k / |\pi^k|)^2$. 
Both the filtering and smoothing procedures (top panels) successfully track the true heterozygosity trajectory (black line), even under this slight model mismatch. See Section \ref{SM:param} for further plots.

Notably, our duality-based filter achieves superior accuracy compared to the Bootstrap Particle Filter (BPF), a leading benchmark in the literature \citep{chopin2020introduction}. While the BPF (bottom left panel) performs reasonably well, it requires expensive simulation of $\PD$-diffusion paths (Eq.~\eqref{eq:latent-transition-hierarchy}) and exact importance weight computations (Eq.~\eqref{likelihood}) for each particle and time step, leading to substantial computational overhead. Furthermore, as seen in Fig.~\ref{fig:entropy2_filter_smooth_mle}, the BPF may occasionally yield posteriors with greater variance and pronounced multimodality, especially at intermediate time points (cf.~also Section \ref{sec: application}).

By contrast, our duality-based approach avoids such burdens, yielding stable, high-quality estimates at a fraction of the computational cost.

Finally, as shown in the bottom right panel, inference based on independent priors (ignoring temporal dependence) may incidentally approximate the true heterozygosity but lacks robustness and statistical adequacy in dynamic or data-sparse regimes. It also precludes interpolation at arbitrary time points, which our model handles naturally and with no additional effort.

\begin{figure}[t!]
    \centering
    \includegraphics[width=.9\linewidth]{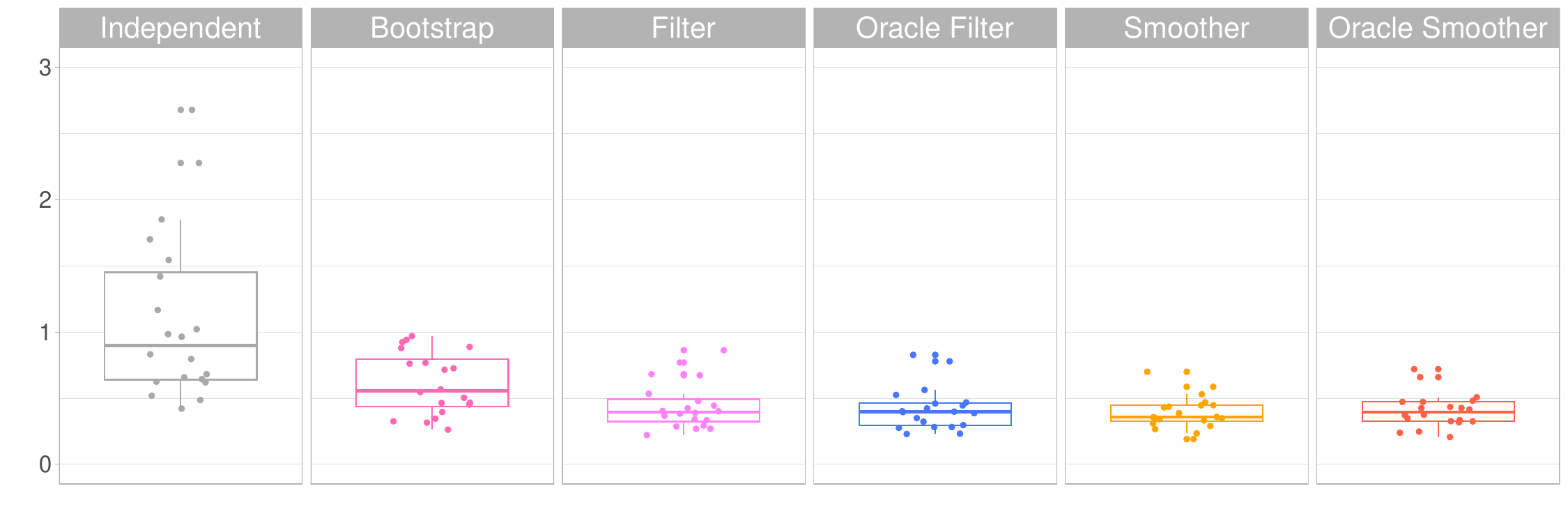}
    \caption{\scriptsize Interval score comparison across five methods. Filtering and smoothing based on MLEs perform comparably to oracle methods.
Bootstrap particle filtering yields less accurate results besides being more demanding computationally.
Independent priors yield lower accuracy due to lack of temporal borrowing.}
    \label{fig:score_vs_ind_filter_smooth_mle}
\end{figure}

To assess accuracy, we compare negatively-oriented interval scores (cf.~\eqref{eq: score}) across six strategies: (i) independent priors at each time point, (ii) bootstrap particle filtering, (iii) filtering with MLE parameters, (iv) \emph{oracle} filtering using true parameters, (v) smoothing with MLE parameters, and (vi) \emph{oracle} smoothing using true parameters.

Fig.~\ref{fig:score_vs_ind_filter_smooth_mle} reports the average scores over 20 replicates of trajectories, each spanning 10 time points with $\Delta_k = 0.01$, partition size 10, and parameters $(\alpha, \theta) = (0.1, 1.0)$. Our filtering and smoothing methods significantly outperform the independent prior baseline and approach the accuracy of an oracle with full knowledge of the latent process. The BPF improves over the independent prior baseline but still falls short of both filtering and smoothing under our duality-based framework, and does so at a substantially higher computational cost.


\subsection{Application to dynamic face-to-face interaction networks}
\label{sec: application}

We apply our methodology to the real-world \textsc{Infectious} dataset\footnote{\url{https://sociopatterns.org/datasets/infectious-exhibition-dynamic-contact-networks/}} \citep{isella2011s}, which records temporal social interactions as dynamic networks. The data were collected at the Science Gallery in Dublin, where participants wore proximity sensors, and an edge was logged whenever a face-to-face interaction lasted more than 20 seconds.

Each day's data yield a time-indexed network in which nodes represent individuals and edges denote interactions. We focus on a single day, June 28, 2009, aggregating the data into 30-minute windows to produce 12 networks. The connected components of each graph induce an unlabelled partition representing clusters of individuals in contact. Fig.~\ref{fig:infectious} in Section~\ref{sec:intro} displays four representative networks, each inducing partitions $(11, 9, 4, 3)$, $(38, 2)$, $(22, 3, 2, 2, 2)$, and $(17, 4, 2, 2, 2)$.

We model the time-varying group structure using a two-parameter Poisson--Dirichlet diffusion for the latent probabilities of cluster sizes. Our inferential target is the heterozygosity functional $H_2(x) := 1 - \sum_{i \ge 1} x_i^2$, which approximates the probability that two randomly chosen individuals belong to different components, serving as a proxy for effective population mixing and epidemic spread.

\begin{figure}[t!]
    \centering
    \includegraphics[width=.9\linewidth]{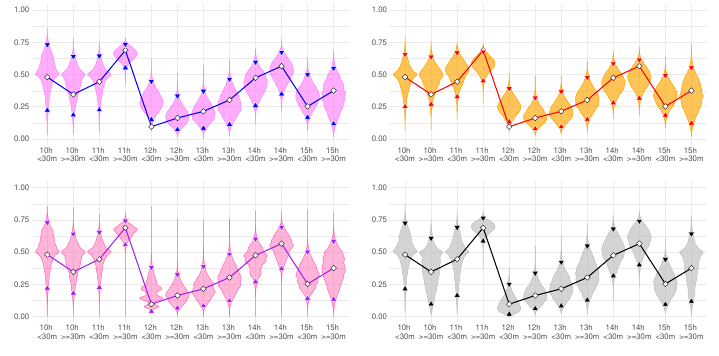}
    \caption{\scriptsize Top: online filtering (left) and offline smoothing (right) of heterozygosity for the \textsc{Infectious} dataset. Bottom: corresponding posteriors from a bootstrap particle filter (left) and independent priors (right). Violin plots: posterior densities; triangles: 95\% credible intervals; white diamonds: observed heterozygosities. All methods use global MLEs $(\hat\alpha, \hat\theta)$ except the independent prior approach, which estimates them separately at each time point.}
    \label{fig:filter_smoothing_infectious}
\end{figure}

Fig.~\ref{fig:filter_smoothing_infectious} reports posterior estimates of $H_2(x)$ across time. Filtering and smoothing are implemented using the marginal MLEs $(\hat \alpha, \hat \theta) = (0, 0.75)$ obtained as in Section~\ref{sec:par estim}. As expected, smoothing provides sharper uncertainty quantification due to backward information flow, while filtering offers a real-time alternative with competitive accuracy.

As in Section \ref{sec:num_exp}, we contrast these results with two commonly used alternatives: the bootstrap particle filter and independent priors fitted separately at each time point. The former produces more dispersed posteriors and occasional multimodality (e.g., at $t_4$), while the latter suffers from inflated uncertainty and instability due to the lack of temporal structure. Further details are provided in Section~\ref{SM:param}.

In terms of computational efficiency, our duality-based filter completes posterior sampling in under two minutes on a standard laptop, while the bootstrap particle filter takes approximately one hour due to path simulation and repeated evaluation of an intractable likelihood \eqref{likelihood}. This highlights the practical advantages of exploiting the discrete dual process for inference in complex time-evolving systems.


\section{Concluding remarks}

We developed an exact and tractable inferential framework for a nonparametric HMM where the latent process is a two-parameter Poisson--Dirichlet diffusion and the observations are unlabelled partitions collected at discrete times. By exploiting Markov duality and combinatorial properties of partition coagulations, we derived recursive updates for both posterior and predictive distributions. These results enabled exact filtering, smoothing, and forecasting algorithms, as well as marginal likelihood-based parameter estimation.

Our approach demonstrates practical scalability and statistical fidelity in modeling time-varying unlabelled partition data. Unlike conventional particle filters, which require simulating high-dimensional latent dynamics and evaluating intractable likelihoods, our method sidesteps both difficulties through structural duality and quasi-conjugacy. The resulting algorithms are substantially faster and more stable, yielding coherent posterior trajectories and sharper uncertainty quantification. As shown through synthetic and real-world experiments, the proposed methodology enables tractable, interpretable, and provably exact inference in settings where the latent structure is infinite-dimensional and only partially observed. These properties make it particularly well suited for applications with computational constraints or unobserved individual-level labels, such as social or genetic data.


\paragraph{Data availability statement.}
The empirical data used in this article are publicly available from the Infectious SocioPatterns repository\footnote{https://sociopatterns.org/datasets/infectious-exhibition-dynamic-contact-networks/} under the CC BY-NC-SA license. The reproducibility materials accompanying the article, including code and cached computational outputs, are available both as online supplementary material and at \url{https://github.com/marcodallapria}.
%


\section*{Acknowledgements}
\addcontentsline{toc}{section}{Acknowledgements}

The authors are grateful to the Associate Editor and the reviewers for their careful reading and helpful suggestions, which led to substantial improvements in the paper's presentation and accompanying reproducibility materials.

MDP and MR acknowledge support of MUR - Prin 2022 - Grant no. 2022CLTYP4, funded by the European Union -- Next Generation EU.

\newpage
\appendix

\def \N {\mathbb{N}}
\def \PP {\mathrm{Pr}}
\def \P {\mathcal{P}}
\def \d {{\rm d}}
\def \H {\mathcal{H}}
\def \F {\mathcal{F}}
\def \Z {\mathbb{Z}}
\def \T {\mathcal{T}}
\def \S {\mathcal{S}}
\def \M {\mathcal{M}}
\def \E {\mathbb{E}}
\def \B {\mathcal{B}}
\def \dualtr {p^{\downarrow}}

\def \ind {{\mathds{1}}}	
\def \iid {\overset{\text{iid}}{\sim}}
\def \simplex {\overline\nabla}
\def \simplexone {\nabla}
\def \tppd {two-parameter Poisson--Dirichlet$\ $}
\def \PD {\mathrm{PD}_{\alpha, \theta}}
\def \pd {\mathrm{PD}}
\def \pr {\Pr_{\alpha,\theta}}
\def \prp {\mathscr{P}_{\alpha,\theta}}
\def \sp {\preccurlyeq}
\def \psf{\mathrm{PSF}_{\alpha, \theta}}
\def \coag{\mathrm{coag}}
\def \Coag{\mathrm{Coag}_{\alpha, \theta}}
\def \csp {C_{\simplexone}([0,\infty))}
\def\crp{\mathcal{C}_{\alpha, \theta}}
\def \Crp{\mathrm{CRP}_{\alpha, \theta}}
\def \paint{\text{Paintbox}}
\def \dual{\mathrm{Dual}}
\def \CRP{\mathrm{CRP}}

\clearpage 
\begin{center}
    \LARGE\bf Supplementary Material
\end{center}
\addcontentsline{toc}{section}{Supplementary Material}
\setcounter{section}{0}
\setcounter{equation}{0}
\setcounter{figure}{0}
\setcounter{table}{0}

\renewcommand{\thesection}{\Alph{section}}
\renewcommand{\theequation}{\thesection.\arabic{equation}}
\renewcommand{\thefigure}{\thesection.\arabic{figure}}
\renewcommand{\thetable}{\thesection.\arabic{table}}


\section{Two-parameter Poisson--Dirichlet diffusions}\label{SM:petrov}

The \tppd diffusion takes values in the closure of the infinite-dimensional ordered simplex, sometimes called Kingman simplex, defined as
\begin{equation}\label{simplex}
\simplex=\bigg\{x\in [0,1]^{\infty}:\ x_{1}\ge x_{2}\ge\ldots\ge 0,  \sum_ {i=1}^{\infty}x_{i}\le 1\bigg\},
\end{equation} 
which is a compact space in the product topology. Its infinitesimal operator is defined, for $\alpha\in[0,1)$ and $\theta\ge-\alpha$.  
\begin{equation}\label{Petrov operator}
A=\frac 12\sum_{i,j=1}^{\infty}x_{i}(\delta_{
ij}-x_{j})\frac {\partial^2}{\partial x_{i}\partial x_{j}}
-\frac{1}{2}\sum_{i=1}^{\infty} (\theta x_{i}+\alpha)\frac {\partial 
}{\partial x_{i}},
\end{equation}
with domain given by the subalgebra of the space $C(\simplex)$ of continuous function on $\simplex$ generated by $\{\varphi_{m}(x),m\in \N\}$, with $\varphi_{1}(x)\equiv 1$ and $\varphi_{m}(x)=\sum_ {i \ge 1} x_{i}^{m}$ for $m\ge2$. See, e.g., \cite{feng2010poisson}, Section 5.5.
The transition function of the diffusion was found in \cite{feng2011functional} through a spectral expansion, and admits the  representation as infinite mixture of conditional \tppd distributions  
\begin{equation}\label{transition}
	p_{t}(x,\cdot)=(d_0(t)+d_1(t))\PD(\cdot) +\sum_{n\ge2} d_{n}(t)\sum_{\pi:|\pi| =n}
P(\pi  |  x)\PD^\pi(\cdot).
\end{equation}
See \cite{griffiths2024dual} for a discussion.
Here $P(\pi  |  x)$ is 
\begin{equation}\label{symmetric polynomial}
P(\pi  |  x):= C(\pi) \sum_{i_1 \neq \dots \neq i_\len} x_{i_1}^{\pi_1} x_{i_2}^{\pi_2} \dots x_{i_\len}^{\pi_\len}.
\end{equation}
and $\PD^\pi$ is the law of a \tppd random probability measure conditional on a partition $\pi$ (cf.~Sec.~\ref{sec:model}).
Furthermore,
\begin{equation} \label{d_n(t)}
	d_n^\theta(t) =
\sum_{k=n}^{\infty}e^{-k(\theta+k-1)t/2}
(-1)^{k-n}
\frac{(2k+\theta-1)(\theta+n)_{(k-1)}}{n!(k-n)!},
\end{equation}
where $a_{(b)}=a(a+1)\cdots(a+b-1)$ is the rising factorial or Pochhammer symbol, with $a_{(0)}:=1$. These $d_{n}(t)$ are the transition probabilities  of a death process on $\Z_{+} = \{0, 1, \dots\}$ that has entrance boundary at $\infty$ and infinitesimal transition rates $n(\theta+n-1)/2$ from $n$ to $n-1$. This process is in fact the \emph{block-counting process} associated to Kingman's coalescent with mutation \citep{tavare1984line}. 

It was showed by \cite{ethier2014property} that, almost surely, the two-parameter diffusion takes values in  
\begin{equation}\label{simplexone}
\nabla=\bigg\{x\in [0,1]^{\infty}:\ x_{1}\ge x_{2}\ge\ldots\ge 0,  \sum_ {i=1}^{\infty}x_{i}= 1\bigg\}
\end{equation} 
at all $t>0$, so that $\simplex\setminus\nabla$ acts as an entrance boundary. Then, if $X(0)\in \nabla$ almost surely, we can take $\nabla$ as the process state space. For this reason in Section \ref{sec:model} we assumed  $X(0)\sim\PD$, which implies $X(0)\in \nabla$ with probability one \citep{pitman2006combinatorial}. All integrals may be taken over $\nabla$, since \(\simplex\setminus\nabla\) is an entrance boundary.


\section{Coagulation of random partitions}\label{SMsec:preliminaries}

\subsection{Two-parameter Poisson--Dirichlet partition structures}\label{SMsec:partition-structures}

An integer \emph{partition} of $n\in\N$ is a nonincreasing vector $\pi=(\pi_1,\ldots,\pi_{\len})$ of positive integers $\pi_{i} \ge \pi_{i+1} > 0$ with $|\pi|:=\sum_{i=1}^{\len}\pi_i=n$, where the \emph{length} $\len$ of $\pi$ counts its parts.
See Figure \ref{fig:diagrams} for examples of Young diagram representation of partitions.
Let $\P_n$ be the set of partitions of $n$ and $\P:=\bigcup_{n\ge1}\P_n$.
For $\pi\in\P$, write $a(\pi):=(a_j(\pi),\,j\ge1)$ where
\begin{equation}\label{allelic multiplicity}
a_j(\pi):=\sum_{i=1}^{\len}\mathbbm{1}\{\pi_i=j\}
\end{equation} 
 is the multiplicity of part size $j$.
We partially order $\P$ by $\pi \sp \lambda$ if the Young diagram of $\pi$ is obtained from that of $\lambda$ by deleting squares; for $\Lambda\subseteq\P$, define the \emph{lower set}
$L(\Lambda):=\{\pi:\ \pi\sp \lambda,\ \lambda\in\Lambda\}$.

A \emph{partition structure} \citep{kingman_representation_1978} is a family $\{P_n:n\in\N\}$ with $P_n$ a probability on $\P_n$ such that $P_n$ is the marginal obtained by removing uniformly at random one square from the Young diagram of $\Pi\sim P_{n+1}$.
A fundamental result by \cite{kingman_random_1975, kingman_representation_1978, kingman_random_1978} says that any partition structure can be represented as a mixture of \emph{paintbox processes},
\begin{equation} \label{kingman representation}
	P(\pi) = \int_{\overline \nabla} P(\pi  |  x) \mathcal{M}(dx)
\end{equation}
for some unique probability measure $\mathcal{M}$ on $\overline \nabla$, where $P(\pi  |  x)$ is the law of the paintbox from $x$ given in Eq. \eqref{symmetric polynomial}, as well as in the main text in Eq. \eqref{likelihood}.

Throughout we use the Ewens--Pitman $(\alpha,\theta)$ partition structure \citep{pitman1995exchangeable,pitman2006combinatorial}.
Its mass function $\psf(\pi)$ is given in the main text at \eqref{PSF}; the associated combinatorial coefficient is
\begin{equation}\label{psf explicit}
C(\pi)\;=\;\binom{|\pi|}{\pi_1,\ldots,\pi_{\len}}\;\frac{1}{a_1(\pi)!\cdots a_{|\pi|}(\pi)!}.
\end{equation}

It has been shown that the representing measure of the Ewens-Pitman Sampling Formula is the two parameter Poisson--Dirichlet distribution (which is supported on $\nabla$), so that Kingman's representation reads
\begin{equation} \label{representation of psf}
	\psf(\pi) = \int_{\nabla} P(\pi  |  x) \PD(\d x);
\end{equation}
see \cite{pitman1995exchangeable, pitman2006combinatorial}.

A random partition $\Pi$ of size $n$ with probability mass function $\{ \psf(\pi)$, $\pi \in \mathcal P_n \}$, is denoted by $\Pi \sim \psf$.
The dependence on $n$ is omitted since $\psf$ defines a partition structure and is therefore consistent across different values of $n$.
Equivalently, $\Pi$ can be generated by the $(\alpha,\theta)$ Chinese restaurant process (CRP): the $i$th item joins an existing block of current size $\pi_j$ with probability $(\pi_j-\alpha)/(\theta+i-1)$, or starts a new block with probability $(\theta+\len \alpha)/(\theta+i-1)$, where $\len$ is the current number of blocks.


\subsection{Coagulation of two-parameter random partitions}\label{SM:coag}

A key ingredient in our framework is the \emph{coagulation} of partitions. Informally, given $\omega,\gamma\in\P$, a coagulation is obtained by summing selected pairs of parts and appending the remainder.

Recall that a \emph{composition} is a partition where the condition $\pi_{i} \ge \pi_{i+1}$ is not enforced, and a \emph{weak composition} allows zeros.
The following makes Definition~\ref{def:coag} in the main text explicit. 

\begin{definition}[Coagulation of partitions]\label{def:coag-supp}
Let $\omega,\gamma\in\P$ and $0\le d\le\min(\len(\omega), \len(\gamma))$.  
For compositions $z,y$ of length $d$ with $a(z)\le a(\omega)$ and $a(y)\le a(\gamma)$, with $a(\cdot)$ as in \eqref{allelic multiplicity}, define
\[
a(\mu)\;=\;a(z+y)\,+\,a(\omega)-a(z)\,+\,a(\gamma)-a(y),
\]
where $\mu\in\P_{|\omega|+|\gamma|}$.  
The coagulation of $\omega$ and $\gamma$ is
\begin{equation}\nonumber\label{coag-sets}
\coag(\omega,\gamma):=\bigcup_{d=0}^{\min(\len(\omega), \len(\gamma))}\ \bigcup_{z,y}\ \coag_{z,y}(\omega,\gamma),
\end{equation}
with the setwise extension $\coag(\Omega,\Gamma):=\bigcup_{\omega\in\Omega,\gamma\in\Gamma}\coag(\omega,\gamma)$.
\end{definition}

See Figure~\ref{fig:diagrams}. Different choices of $z,y$ may yield the same coagulation. This operation is crucial for describing products of likelihoods as in~\eqref{symmetric polynomial}.

\begin{proposition}\label{prop:prod-cond-SM}
Let $x\in\nabla$, $\omega\in\P_n$, $\gamma\in\P_m$, and $P(\cdot|x)$ as in \eqref{symmetric polynomial}. Then
\begin{equation}\label{prod-cond}
P(\omega  |  x) P(\gamma  |  x) = \sum_{\mu\in\coag(\omega,\gamma)} \H(\omega,\gamma  |  \mu) P(\mu  |  x),
\end{equation}
where
\[
\H(\omega,\gamma  |  \mu)=\binom{n+m}{n,m}^{-1}\sum \prod_{j=1}^{\len(\mu)}\binom{\mu_j}{z_j,y_j},
\]
and the sum runs over weak compositions $z,y$ of $n,m$ consistent with $\omega,\gamma$.
\end{proposition}

Proposition~\ref{prop:prod-cond-SM} underpins quasi-conjugacy: while products of likelihoods exit the family, they remain finite linear combinations of the same kernels, corresponding to coagulated partitions.

\begin{proposition}\label{prop:joint-crp}
For $\omega\in\P_n,\ \gamma\in\P_m,\ \mu\in\coag(\omega,\gamma)$,
\[
\Pr(\Pi_{1:n+m}=\mu,\ \Pi_{n+1:n+m}=\gamma |  \Pi_{1:n}=\omega)
=\H(\omega,\gamma  |  \mu)\,\frac{\psf(\mu)}{\psf(\omega)}.
\]
\end{proposition}
Integrating over $\mu$ in the statement of Proposition \ref{prop:joint-crp} yields
\begin{equation}\label{gamma-given-omega}
\psf^{\omega}(\gamma) := \Pr(\Pi_{n+1:n+m}=\gamma |  \Pi_{1:n}=\omega)
=\sum_{\mu\in\coag(\omega,\gamma)} \H(\omega,\gamma|\mu)\,\frac{\psf(\mu)}{\psf(\omega)},
\end{equation}
the conditional Pitman Sampling Formula.
Moreover,
\begin{equation}\label{mu-given-omega-gamma}
    \Pr( \Pi_{1:n+m} = \mu  |  \Pi_{1:n} = \omega , \Pi_{n+1:n+m} = \gamma ) \propto \H (\omega, \gamma  |  \mu) \psf(\mu), \quad \mu \in \coag(\omega, \gamma)
\end{equation}
and we will write $\Pi_{1:n+m}  |  \Pi_{1:n} = \omega , \Pi_{n+1:n+m} = \gamma \sim \Coag(\omega, \gamma)$, the \emph{coagulation kernel}.
Consider now $\Pi_{1:n} \sim v_{\omega}$ where $\{v_{\omega}; \omega \in \Omega \subset \P\}$ is \emph{any} probability mass function on a finite subset of $\P$.
Assume that $\Pi_{n+1:n+m}  |  \Pi_{1:n} = \omega \sim \psf^{\omega}$, i.e., given the current configuration $\omega$ of the restaurant the next $m$ customers join $\omega$ according to a Chinese Restaurant Process.
Then, conditionally on $\Pi_{n+1:n+m}=\pi$, the distribution of $\Pi_{1:n}$ is updated as
\[
\hat v_\omega\;\propto\; v_\omega\,\psf^{\omega}(\pi),
\]
and the joint law of $\Pi_{1:n+m}$ (the full configuration of the restaurant) is a mixture of coagulation kernels:
\begin{equation}\label{mix-coag}
\Pi_{1:n+m}  |  \Pi_{n+1:n+m}=\pi\;\sim\;\sum_{\omega}\hat v_\omega\,\Coag(\omega,\pi).
\end{equation}
This sequential structure will be the basis for filtering $\PD$ diffusions given partition-valued observations.


\subsection{A CRP with loss of particles as dual process}\label{SMsubsec:dual}

Two Markov processes $X=(X(t);t\ge0)$ and $D=(D(t);t\ge0)$ with state spaces $\mathcal{X},\mathcal{D}$ are said to be \emph{dual} with respect to a bounded measurable function $H:\mathcal{X}\times\mathcal{D}\to\mathbb{R}$ if
\[
\E\!\left[H(X(t),d) |  X(0)=x\right]
=\E\!\left[H(x,D(t)) |  D(0)=d\right].
\]
See \citet{jansen2014notion} for background.

\citet{griffiths2024dual} showed that the dual process of the \tppd diffusion is a continuous-time Markov chain on partitions $\P$, with transition kernel
\begin{equation}\label{dual transition}
\dualtr_{\lambda,\omega}(t):=\Pr(D(t)=\omega  |  D(0)=\lambda)
= \H(\omega  |  \lambda)\, d_{|\lambda|,|\omega|}^\theta(t),\quad \omega \in L(\lambda),
\end{equation}
where $H(x,\omega)=P(\omega  |  x) / \psf(\omega)$, $\H(\omega | \lambda):=\sum_{\gamma}\H(\omega,\gamma | \lambda)$ with $\H(\omega,\gamma | \lambda)$ as in  Proposition~\ref{prop:prod-cond-SM}, and 
$d_{n,m}^\theta(t):=\Pr(D(t)=m |  D(0)=n)$ is the transition probability of the block-counting process of Kingman's coalescent. Thus the dual is a \emph{pure-death process} on integer partitions: from state $\lambda$, jumps occur at rate $|\lambda|(|\lambda|+\theta-1)/2$, and a uniformly chosen square from the Young diagram of $\lambda$ is deleted. A random partition with law~\eqref{dual transition} will be denoted $\Pi\sim\dualtr_{\lambda,\bullet}(t)$.

As in~\eqref{mix-coag}, mixtures of dual processes arise naturally. If the initial state is random with distribution $\{w_\lambda,\ \lambda\in\P\}$, then
\[
\Pi \sim \sum_{\lambda} w_\lambda\, \dualtr_{\lambda,\bullet}(t)
\]
is the distribution of the dual process started from this mixture.  

This dual representation is central to our methodology: it enables tractable filtering of a \tppd hidden Markov model, with the random loss of particles in the dual mirroring the contraction of information through time in the latent diffusion.


\section{Details and proofs on filtering and smoothing}\label{SM:filtering}

\subsection{Online estimation: filtering and prediction}\label{SMsec:prediction}

We introduce two operators mapping probability measures on $\simplexone$ into themselves.  

The first is the \emph{propagation operator}
\begin{equation}\label{propagation operator}
\varphi_t(\nu)(\d x')=\nu \T_t(\d x')
:=\int_\nabla \nu(\d x)\,p_t(x,\d x'),\qquad
\T_t f(x):=\int_\nabla f(x')p_t(x,\d x'),
\end{equation}
where $\{ \T_t, t\ge0\}$ is the semigroup of the $\PD$ diffusion with transition kernel $p_t(x,\d x')$.
While $p_t$ admits only an infinite mixture expansion (cf.~\eqref{transition}), we will see that $\varphi_t(\nu)$ is finite whenever $\nu$ is.

The second operator, denoted $\phi_\pi$, acts as a Bayesian update: if $X(t_k)\sim\nu$, then $\phi_\pi(\nu)$ is the law of $X(t_k) |  \Pi^k=\pi$.

Starting from $X(t_0)\sim \PD$, the filtering recursion is
\begin{equation}\label{recursion}
\nu_{0|0}:=\phi_{\Pi^0}(\PD),\qquad
\nu_{k|0:k}=\phi_{\Pi^k}\big(\varphi_{\Delta_k}(\nu_{k-1|0:k-1})\big),\quad
\Delta_k:=t_k-t_{k-1}.
\end{equation}

\paragraph{Propagation.}
Suppose $\nu_{k-1|0:k-1}=\sum_{\lambda\in\Lambda_{0:k-1}}w_\lambda \PD^\lambda$.
Propagation over $\Delta_k$ yields:

\begin{theorem}[Propagation]\label{thm:propagation}
The predictive law $\nu_{k|0:k-1}$ of $X(t_k) |  \Pi^{0:k-1}$ belongs to $\F$:
\begin{equation}\label{propagation}
X(t_k) |  \lambda(t_k)\sim \PD^{\lambda(t_k)},\qquad
\lambda(t_k)\sim \sum_{\lambda\in\Lambda_{0:k-1}}w_\lambda\,\dualtr_{\lambda,\bullet}(\Delta_k),
\end{equation}
where $\dualtr_{\lambda,\bullet}(\Delta_k)$ are the dual transition probabilities \eqref{dual transition}.  
Equivalently,
\[
\nu_{k|0:k-1}=\sum_{\omega\in L(\Lambda_{0:k-1})} v_\omega\,\PD^\omega,\qquad
v_\omega=\sum_{\lambda\in\Lambda_{0:k-1}} w_\lambda\,\dualtr_{\lambda,\omega}(\Delta_k).
\]
\end{theorem}

This shows that although the transition density of the $\PD$ diffusion is an doubly infinite sum (cf.~\eqref{transition} and \eqref{d_n(t)}), its effect on finite mixtures is finite.  
Sampling proceeds as follows:
\begin{enumerate}
\item draw $\lambda(t_{k-1})\sim\{w_\lambda; \lambda \in \Lambda_{0:k-1}\}$,
\item draw $\lambda(t_k)\sim \dualtr_{\lambda(t_{k-1}),\bullet}(\Delta_k)$,
\item draw $X(t_k)\sim \PD^{\lambda(t_k)}$.
\end{enumerate}

The following corollary establishes the convergence of the propagated law to stationarity.
\begin{corollary}\label{cor: convergence to stationarity}
Let $\varphi_{t}(\nu_{N|0:N})(\cdot):=\Pr(X(t_{N}+t)\in \cdot  |  \Pi^{0:N})$. Then 
\begin{equation}\nonumber
\varphi_{t}(\nu_{N|0:N}) \rightarrow \PD,
\end{equation} 
in total variation, as $t\rightarrow \infty$.
\end{corollary}

\paragraph{Update.}
Given $\Pi^k=\pi^k$, we condition $\nu_{k|0:k-1}$ to obtain:

\begin{theorem}[Update]\label{thm:update}
If $\nu_{k|0:k-1}=\sum_{\omega\in L(\Lambda_{0:k-1})} v_\omega \PD^\omega$, then
\begin{equation}\label{updated distribution described}
X(t_k) |  \lambda(t_k)\sim \PD^{\lambda(t_k)},\qquad
\lambda(t_k) |  \pi^k \sim \sum_{\omega\in L(\Lambda_{0:k-1})}\hat v_\omega \Coag(\omega,\pi^k),
\end{equation}
with $\hat v_\omega\propto v_\omega \psf^{\omega}(\pi^{k})$, where $\psf^{\omega}$ is as in \eqref{conditional CRP} and $\Coag$ as in \eqref{mu-given-omega-gamma}.  
Equivalently,
\[
\nu_{k|0:k}=\sum_{\lambda\in\Lambda_{0:k}} w_\lambda \PD^\lambda,\qquad
w_\lambda \propto \sum_{\omega\in L(\Lambda_{0:k-1})}\hat v_\omega\,\Pr(\lambda |  \omega,\pi^k),
\]
with $\Lambda_{0:k}=\coag(L(\Lambda_{0:k-1}),\pi^k)$, and $\Pr(\lambda  |  \omega, \pi^k)$ is shorthand for the distribution on $\coag(\omega, \pi^{k})$ as in Eq. \eqref{mu-given-omega-gamma}.
\end{theorem}

The result follows because $P(\pi^k  |  x)$ can be factorized kernel-by-kernel in $\nu_{k|0:k-1}$, and Proposition~\ref{prop:prod-cond-SM} expresses the resulting product as a mixture over coagulations.  

\paragraph{Filtering algorithm.}
Theorems~\ref{thm:propagation} and~\ref{thm:update} yield an exact recursive filtering scheme for the HMM~\eqref{eq:hmm-structure}, summarized in Algorithm~\ref{alg: generic filtering}.  
Practical aspects, including approximation strategies, are discussed in Section~\ref{SM:alg}.


\subsection{Offline estimation: smoothing and data forecasting}\label{SMsec:smoothing}
After observing $\Pi^{0:N}$, we (i) refine state estimates at past observation times and (ii) generate future or interpolated data. For $k\le N$, define the \emph{marginal smoother}
\begin{equation}\label{marginal smoother}
\nu_{k|0:N}(A):=\Pr \big(X(t_k)\in A  |  \Pi^{0:N}\big).
\end{equation}

\paragraph{Backward information (cost-to-go).}
Define 
\[
\Pr(\Pi^{k:N}  |  X(t_{k-1})=x)
\]
as the likelihood of the remaining partitions $\Pi^{k:N}$ given the state at $t_{k-1}$. This quantity, also known as the \emph{cost-to-go function} or \emph{backward information filter}, admits a general recursive representation.  
For intuition, in the simplest case $k=N$ one obtains
\begin{equation}\label{last cost to go}
\Pr(\Pi^N=\pi |  X(t_{N-1})=x)
=\psf(\pi)\sum_{\lambda\in L(\pi)} h_\lambda(\Delta_N)\,\frac{P(\lambda |  x)}{\psf(\lambda)},
\qquad \pi\in\P_{n_N},
\end{equation}
where $h_\lambda(\Delta_N)$ is a distribution on $L(\pi)$. The sum is finite (cf. Fig.~\ref{fig:lower set}).

The general form is as follows.

\begin{proposition}[Cost-to-go]\label{thm:cost to go}
Let $\Pi^{0:N}=\pi^{0:N}$ be the observed partitions. Then
\begin{align}\label{cost to go in thm}
\Pr(\Pi^k=\pi^k,\dots,\Pi^N=\pi^N  |  X(t_{k-1})=x)
&= \prod_{j=k}^N g(\pi^j |  \pi^{j+1:N})\,
\sum_{\omega\in \Omega^{k:N}} h_\omega(\Delta_k)\,\frac{P(\omega |  x)}{\psf(\omega)},
\end{align}
where
\[
g(\pi^N)=\psf(\pi^N),\quad
h_\omega(\Delta_N)=\dualtr_{\pi^N,\omega}(\Delta_N),\ \ \omega\in\Omega^{N:N},
\]
and for $k=1,\ldots,N-1$,
\[
g(\pi^k |  \pi^{k+1:N})
=\sum_{\omega\in \Omega^{k+1:N}} h_\omega(\Delta_{k+1}) \psf^{\omega}(\pi^{k}).
\]
The backward weights $h_\omega(\Delta_k)$ are computed recursively via
\[
h_\omega(\Delta_k)
=\sum_{\mu\in \coag(\Omega^{k+1:N},\pi^k)}
\left[\sum_{\omega'\in \Omega^{k+1:N}}
\frac{h_{\omega'}(\Delta_{k+1}) \psf^{\omega'}(\pi^{k}) }
{g(\pi^k |  \pi^{k+1:N})}\,
\Pr(\mu |  \omega',\pi^k)\right]\dualtr_{\mu,\omega}(\Delta_k),
\]
with $\psf^{\omega'}$ as in~\eqref{gamma-given-omega} and 
$\Pr(\mu |  \omega',\pi^k)$ as in~\eqref{mu-given-omega-gamma}.
\end{proposition}

\paragraph{Forward and backward supports.}
To combine forward and backward information, we define two deterministic sets of partitions.  
The forward support $\Lambda_{0:k}$ is the set of partitions that carry positive mass after filtering up to time $t_k$ (Theorems~\ref{thm:propagation}, \ref{thm:update}).  
The backward supports are defined recursively from the observed data:
\begin{equation}\label{big lambdas}
\Omega^{N:N}:=L(\pi^N),
\qquad
\Omega^{k:N}:=L\!\big(\coag(\Omega^{k+1:N},\pi^k)\big),\quad k=N-1,\dots,0,
\end{equation}
with $L(\cdot)$ and $\coag(\cdot,\cdot)$ as in Sections~\ref{SMsec:partition-structures} and~\ref{SM:coag}.  
Intuitively, $\Lambda_{0:k}$ encodes forward uncertainty up to $t_k$, while $\Omega^{k:N}$ encodes backward constraints imposed by future data.

Having both components, we can now combine them to obtain the full smoothing distribution.

\begin{theorem}[Smoothing]\label{thm:smoothing}
For $k=0,\dots,N$, the smoothed law of $X(t_k) |  \Pi^{0:N}$ belongs to $\F$ and can be written as
\begin{equation}\label{smoothing distribution}
X(t_k) |  \lambda(t)\sim \PD^{\lambda(t)},\qquad
\lambda(t) |  \Pi^{0:N}\sim \sum_{(\lambda,\omega)\in \Lambda_{0:k}\times \Omega^{k:N}}
w_{\lambda,\omega}\,\Coag(\lambda,\omega),
\end{equation}
where
\begin{equation}\label{SM:smoothing weights described}
w_{\lambda,\omega}\ \propto\ 
v_\lambda\,
\frac{\Pr\!\big(\Pi_{1:|\lambda|}=\lambda,\ \Pi_{|\lambda|+1:|\lambda|+|\omega|}=\omega\big)}{\psf(\lambda)\,\psf(\omega)}
\,h_\omega,
\end{equation}
$v_\lambda$ are the forward weights on $\Lambda_{0:k}$ (Theorem~\ref{thm:propagation}), and $h_\omega$ are the backward weights on $\Omega^{k:N}$ from Proposition~\ref{thm:cost to go}. See Section~\ref{SM:proofs} for the proof and Section~\ref{SM:mixw} for further discussion.
\end{theorem}

Algorithm~\ref{alg: smoothing} evaluates $\nu_{k|0:N}$ by combining the forward mixture at $t_k$ with the backward information summarized by $\Omega^{k:N}$ via coagulation.

\paragraph{Interpolation between observation times.}
The same structure yields $\Pr(X(t_k+\delta)\in\cdot |  \Pi^{0:N})$ for $0<\delta<\Delta_{k+1}$.

\begin{corollary}[Interpolation]\label{cor:interpolation}
For $t_k+\delta\in(t_k,t_{k+1})$, $\Pr(X(t_k+\delta)\in\d x |  \Pi^{0:N})\in\F$ and
\begin{equation}\label{smoothing distribution arb times}
\Pr(X(t_k+\delta)\in\d x |  \Pi^{0:N})
=\sum_{\mu\in \coag\big(L(\Lambda_{0:k}),\ \Omega^{k:N}\big)}
s_\mu(\delta,\Delta_{k+1})\,\PD^\mu(\d x),
\end{equation}
with (up to normalization)
\begin{equation}\label{interpolation weights described}
s_\mu(\delta,\Delta_{k+1})\ \propto\ 
\sum_{\eta\in L(\Lambda_{0:k})}\sum_{\omega\in \Omega^{k:N}}
w_{\eta,\omega}(\delta,\Delta_{k+1})\,\Pr(\mu |  \eta,\omega),
\end{equation}
\begin{equation}
w_{\eta,\omega}(\delta,\Delta_{k+1})
\propto w_\eta(\delta)\,
\frac{\Pr\!\big(\Pi_{1:|\eta|}=\eta,\ \Pi_{|\eta|+1:|\eta|+|\omega|}=\omega\big)}{\psf(\eta)\psf(\omega)}
\,h_\omega(\Delta_{k+1}-\delta),
\end{equation}
where $w_\eta(\delta)$ are the forward propagation weights over $\delta$ (Theorem~\ref{thm:propagation}) and $h_\omega(\cdot)$ are cost-to-go weights (Proposition~\ref{thm:cost to go}).
\end{corollary}

For $t>t_N$, use the forecast $\varphi_{t-t_N}(\nu_{N|0:N})$ (Theorem~\ref{thm:propagation} and Corollary~\ref{cor: convergence to stationarity}).

\paragraph{Forecasting new partition data.}
We can also simulate future (or interpolated) partitions $\Pi'(t)$ conditional on $\Pi^{0:N}$.

\begin{corollary}[Forecasting partitions]\label{cor: forecasting}
For any $t\ge0$ under \eqref{eq:hmm-structure},
\begin{equation}\label{eq: forecasting}
\Pi'(t) |  \lambda(t)\ \sim\ \psf^{\lambda(t)},
\qquad
\lambda(t)\ \sim\ \{w_\lambda; \lambda \in \Lambda_{t} \},
\end{equation}
where $\{w_\lambda; \lambda \in \Lambda_{t}\}$ are the mixture weights for $\Pr(X(t)\in\d x |  \Pi^{0:N})$, obtained from:
Theorem~\ref{thm:propagation} if $t>t_N$; 
Theorem~\ref{thm:update} if $t=t_N$; 
Theorem~\ref{thm:smoothing} if $t=t_k< t_N$; 
or Corollary~\ref{cor:interpolation} if $t\in(t_k,t_{k+1})$.
\end{corollary}

Algorithm~\ref{alg: forecasting} then samples partitions conditionally on $\pi^{0:N}$.
As in Section \ref{SMsubsec:dual}, Equation~\eqref{eq: forecasting} can be seen as a temporally informed Chinese Restaurant Process, where temporal dependence is encoded by the weights $\{w_\lambda; \lambda \in \Lambda_{t}\}$; the overall clustering structure is unknown within the finite set of plausible coagulations


\subsection{Proofs}\label{SM:proofs}

\emph{Proof of Proposition \ref{prop:prod-cond-SM}.}
We give here a direct probabilistic proof which highlights the role of the random sampling. See also \cite{feng2011functional, feng2010poisson} for previous discussion of similar results.
Conditionally on $x \in \nabla$, $\Pi_{1:n}$ and $\Pi_{n+1:n+m}$ are respectively the clusterings of the first $n$ and of the next $m$ samples of Kingman's paintbox process from $x$.
Then by definiton of the paintbox process the joint probability factorizes as
\begin{equation}\nonumber
\PP_x(\Pi_{1:n} = \omega, \Pi_{n+1:n+m} = \gamma ) = \PP_{x}(\Pi_{1:n} = \omega) \PP_{x}(\Pi_{n+1:n+m} = \gamma) = P(\omega  |  x) P(\gamma  |  x).
\end{equation} 
Moreover, it is plain to see that by definition of the paintbox process, conditionally on the partition $\Pi_{1:n+m}$ obtained by all $n+m$ samples, the probability
$$\PP_x(\Pi_{1:n} = \omega, \Pi_{n+1:n+m} = \gamma  |  \Pi_{1:n+m} = \mu)$$
is in fact independent of $x$ and coincides with the sum of joint hypergeometric probabilities in \eqref{prop:prod-cond-SM}.
Then
\begin{align}
    P(\omega  |  x) P(\gamma  |  x)
    &= \PP_x(\Pi_{1:n} = \omega, \Pi_{n+1:n+m}= \gamma ) \\
    &= \sum_{\mu \in\P_{n+m}} \PP_x( \Pi_{1:n} = \omega, \Pi_{n+1:n+m} = \gamma, \Pi_{1:n+m} = \mu ) \\
    &= \sum_{\mu \in\P_{n+m}} \PP_x(\Pi_{1:n+m} = \mu) \PP_x(\Pi_{1:n} = \omega, \Pi_{n+1:n+m}= \gamma  |  \Pi_{1:n+m} = \mu) \\
    &= \sum_{\mu \in\P_{n+m}} 
 \PP_x(\Pi_{1:n+m} = \mu) \H (\omega, \gamma  |  \mu)
\end{align}
but $\H (\omega, \gamma  |  \mu) = 0$ if $\mu \notin \coag(\omega, \gamma)$, i.e., if $\mu$ cannot be obtained as a coagulation of $\omega,\gamma$, yielding the result. \qed

\emph{Proof of Proposition \ref{prop:joint-crp}.}
By virtue of Kingman's representation \ref{kingman representation}, we known that the marginal law on exchangeable partitions can be represented as a mixture of paintboxes.
In particular, as in \ref{representation of psf}, the Chinese Restaurant Process is obtained by integrating out $X \sim \PD$ from the law of the paintbox of $X$.
Then
\begin{equation}\label{conditional PSF}
\begin{aligned}
\Pr(\Pi_{n+1:n+m}=\gamma |  \Pi_{1:n}=\omega)
=&\,\int_\nabla P(\gamma  |  x)\,\PD^\omega(\d x),\\
\PD^\omega(\d x)=&\,\frac{P(\omega  |  x)\PD(\d x)}{\psf(\omega)}.
\end{aligned}
\end{equation}
See also \cite{griffiths2024dual}.
Using \eqref{conditional PSF}, the left-hand side of Proposition \ref{prop:joint-crp} can be written as a mixture,
\begin{equation}\nonumber
\begin{aligned}
    \Pr(&\,\Pi_{1:n+m} = \mu, \Pi_{n+1:n+m} = \gamma  |  \Pi_{1:n} = \omega ) \\
    =&\, \int_\nabla \PP_x(\Pi_{1:n+m} = \mu, \Pi_{n+1:n+m} = \gamma   |  \Pi_{1:n} = \omega) \PD^\omega(\d x).
\end{aligned}
\end{equation}
where $\PP_x$ denotes the law of the paintbox from $x \in \nabla$; cf. \eqref{symmetric polynomial}.
Using \eqref{conditional PSF}, the previous equals
\begin{equation}\label{step in lemma}
\int_\nabla \PP_x( \Pi_{1:n+m} = \mu, \Pi_{n+1:n+m} = \gamma, \Pi_{1:n} = \omega ) \frac{\PD(\d x )}{\psf(\omega)} 
\end{equation}
since by definition $P(\omega  |  x)=\PP_x(\Pi_{1:n} = \omega)$.
Now, conditionally on $\Pi_{1:n+m}$ the subpartitions $\Pi_{1:n}$ and $\Pi_{n+1:n+m}$ are independent of $x$, (cf.~proof of Proposition \ref{prop:prod-cond-SM}) and their conditional joint law is the hypergeometric in Proposition \ref{prop:prod-cond-SM}.
Hence we have
\begin{equation}\nonumber
\begin{aligned}
    &\PP_x( \Pi_{1:n+m} = \mu, \Pi_{n+1:n+m} = \gamma, \Pi_{1:n} = \omega ) \\
    &= \PP_x( \Pi_{1:n+m} = \mu ) \PP_x( \Pi_{n+1:n+m} = \gamma, \Pi_{1:n} = \omega  |  \Pi_{1:n+m} = \mu ) \\
    &= P(\mu  |  x) \H (\omega, \gamma  |  \mu),
\end{aligned}
\end{equation}
from which \eqref{step in lemma} becomes
\begin{equation}\nonumber
\frac{\H (\omega, \gamma  |  \mu) }{\psf(\omega)} \int_\nabla P(\mu  |  x) \PD(\d x ) = \H (\omega, \gamma  |  \mu) \frac{\psf(\mu) }{\psf(\omega)},
\end{equation}
which follows once again from \ref{representation of psf}.\qed

\emph{Proof of Theorem \ref{thm:propagation}}.
From \eqref{propagation operator} we have
\begin{equation}\label{propagation generic integral}
    \begin{aligned}
\nu_{k|0:k-1}(\d x) 
=&\,\varphi_{\Delta_k}(\nu_{k-1|0:k-1})(\d x)
=\int_\nabla \nu_{k-1|0:k-1}(\d z) p_{\Delta_{k}}(z, \d x )\\
=&\, \sum_{\lambda \in \Lambda} w_\lambda 
\int_\nabla \PD^{\lambda}(\d z)
p_{\Delta_{k}}(z,\d x ).
    \end{aligned}
\end{equation} 
Using now \eqref{conditional PSF}, the previous yields
\begin{equation}\nonumber
\begin{aligned}
\sum_{\lambda \in \Lambda} 
w_\lambda
\int_\nabla 
\frac{ P(\lambda  |  z) }{\psf(\lambda)} \PD(\d z)
 p_{\Delta_{k}}(z, \d x ).
\end{aligned}
\end{equation} 
Since the signal is reversible with respect to $\PD$ (cf.~\cite{petrov2009}, p.~280), using the detailed balance condition we obtain
    \begin{align}\label{propagation pre duality}
        \PD(\d x ) \sum_{\lambda \in \Lambda} w_\lambda
         \int_\nabla \frac{ P(\lambda  |  z)}{\psf(\lambda)} p_{\Delta_{k}}(x, \d z).
\end{align}
Due to the duality between the $\PD$ diffusion and the death process described in Section \ref{SMsubsec:dual}, \eqref{propagation pre duality} becomes
\begin{equation}\nonumber
\begin{aligned}
\PD(\d x ) \sum_{\lambda \in \Lambda} w_\lambda 
\E\big(H(x,D_{\Delta_k})  |  D_{0}=\lambda\big)
\end{aligned}
\end{equation} 
and since $D(t)$ is a death process on $\P$, we obtain
\begin{equation}\nonumber
\begin{aligned}
\PD(\d x ) \sum_{\lambda \in \Lambda} w_\lambda \sum_{\omega \in L(\lambda)} \frac{P(\omega  |  x)}{ \psf(\omega) }\dualtr_{\lambda, \omega}(\Delta_{k})
\end{aligned}
\end{equation} 
which upon a rearrangement and using again \eqref{conditional PSF} yields the result.
\qed

\emph{Proof of Corollary \ref{cor: convergence to stationarity}}.
For any measurable $A \subset \simplexone$, from \eqref{propagation} with $k=N$  and $t=\Delta_{N+1}$, we have
\begin{equation}\nonumber
\begin{aligned}
|\varphi_{t}(\nu_{N|0:N})(A)&-\PD(A)|
\le 
\sum_{\eta \in L(\Lambda)} v_{\eta}(t) \bigg|\PD^\eta(A) - \PD(A)\bigg|\\
\le&\,
v_{\emptyset}(t) \bigg|\PD(A)-\PD(A)\bigg|
+v_{(1)}(t) \bigg|\PD^{(1)}(A) - \PD(A)\bigg|\\
&\,+\sum_{\eta \in L(\Lambda_{0:N}):|\eta|>1} v_{\eta}(t) \bigg|\PD^\eta(A)-\PD(A)\bigg|.
\end{aligned}
\end{equation} 
Now, since 
\begin{equation}\nonumber
v_{\eta}(t)
=\sum_{\lambda \in \Lambda_{0:N}} w_\lambda \H(\eta  |  \lambda)d_{|\lambda|,|\eta|}^\theta(t)
\end{equation} 
the fact that $d_{|\lambda|,|\eta|}^\theta(t)$ is the transition probability of a death-process in $\Z_{+}$ with absorption in $(1)$ implies that $v_{\eta}(t)\rightarrow 0$ as $t\rightarrow \infty$ for all $\eta$ such that $|\eta|>1$. The result is now implied by the fact that $\PD^{(1)} \overset{d}{=}\PD$ (see Remark \ref{singleton-exception}).\qed

\emph{Proof of Theorem \ref{thm:update}}.
An application of Bayes' theorem yields
\begin{equation}
    \nu_{k|0:k}(\d x)\propto P(\pi^k  |  x) \nu_{k|0:k-1}(\d x),
\end{equation}
so that substituting $\nu_{k|0:k-1}(\d x)$ as in Theorem \ref{thm:propagation} and using \eqref{conditional PSF}, we obtain
\begin{equation}\nonumber
    \begin{aligned}
        \sum_{\omega \in L(\Lambda_{0:k-1})} v_\omega P(\pi^k  |  x) P(\omega  |  x) \frac{\PD(\d x )}{ \psf(\omega) }.
    \end{aligned}
\end{equation} 
Using now Propositions \ref{prop:prod-cond-SM}  and \ref{prop:joint-crp} leads to
\begin{align}
    \sum_{\omega \in L(\Lambda_{0:k-1}) }&\,v_\omega
     \sum_{\mu \in \coag(\omega, \pi^k )}  \H (\omega, \pi^k  |  \mu) P(\mu  |  x) \frac{\PD(\d x )}{ \psf(\omega) } \\
    =&\, \sum_{\omega \in L(\Lambda_{0:k-1}) }v_\omega
     \sum_{\mu \in \coag(\omega, \pi^k )} \H (\omega, \pi^k  |  \mu) \frac{\psf(\mu)}{ \psf(\omega) } \PD^\mu(\d x) \\
    =&\, \sum_{\omega \in L(\Lambda_{0:k-1}) } v_\omega 
    \sum_{\mu \in \coag(\omega, \pi^k )} \Pr(\mu, \pi^k  |  \omega) \PD^\mu(\d x).
\end{align}
Since the two sums are finite, we can normalize the last expression by
\begin{equation}\nonumber
    \begin{aligned}
        \sum_{\omega \in L(\Lambda_{0:k-1}) }&\, v_\omega 
        \sum_{\mu \in \coag(\omega, \pi^k)} \Pr(\mu, \pi^k | \omega)
        =\sum_{\omega \in L(\Lambda_{0:k-1}) } v_\omega 
        \psf^{\omega}(\pi^{k})
    \end{aligned}
\end{equation} 
and multiply and divide by $\psf^{\omega}(\pi^{k})$ to obtain
\begin{equation}\nonumber
    \sum_{\omega \in L(\Lambda_{0:k-1}) }
    \frac{ v_\omega \psf^{\omega}(\pi^{k}) }
    {\sum_{\omega'} v_{\omega'} \psf^{\omega'}(\pi^{k}) }
    \sum_{\mu \in \coag(\omega, \pi^k)}
     \Pr(\mu  |  \omega, \pi^k) 
    \PD^\mu(\d x). 
\end{equation}
A rearrangement leads to
\begin{equation}\nonumber\label{filtering highlighting latent variables}
    \sum_{\mu \in \coag(L(\Lambda_{0:k-1}), \pi^k)}
    \left[
    \sum_{\omega \in L(\Lambda_{0:k-1}) }
    \hat{v}_{\omega}
     \Pr(\mu  |  \omega, \pi^k)
    \right]
    \PD^\mu(\d x).
\end{equation}  
\qed

\emph{Proof of Proposition \ref{thm:cost to go}}.
Let us avoid to write the random partitions $\Pi^{k:N}$ in $\Pr$ to ease the notation.
We proceed by induction on $k$.
First note that
\begin{align}
    \Pr(\pi^N   |   X(t_{N-1}) = x) 
    =&\,  \int_\nabla P(\pi^N  |  z)p_{\Delta_{N}}(x, \d z) \\
        =&\, \psf(\pi^N) \int_\nabla \frac{P(\pi^N  |  z)}{\psf(\pi^N)} p_{\Delta_{N}}(x, \d z) \\
    =&\, \psf(\pi^N) \sum_{\omega \in \Omega^{N:N} }  \dualtr_{\pi^N, \omega}(\Delta_{N}) \frac{P(\omega  |  x)}{\psf(\omega)},
\end{align}
where in the last identity we have used the duality (recall Section \ref{SMsubsec:dual}).
Now, since
\begin{align}\label{costo to go intermediate recursion}
\Pr(\pi^{k:N}  |  X(t_{k-1}) = x)
    &= \int_\nabla \Pr(\pi^{k:N}  |  X(t_{k}) = z) p_{\Delta_{k}}(x, \d z) \\
    &= \int_\nabla P(\pi^k  |  z) \Pr(\pi^{k+1:N}  |  X(t_k) = z) p_{\Delta_{k}}(x, \d z),
\end{align}
if we now assume 
\begin{equation}\label{induction hypothesis}
    \Pr(\pi^{k+1:N}   |   X(t_k) = x) = \prod_{j = k+1}^N g(\pi^j  |  \pi^{j+1:N} ) \sum_{ \omega \in \Omega^{k+1:N} } h_\omega(\Delta_{k+1}) \frac{P(\omega  |  x)}{ \psf(\omega)},
\end{equation}
for some functions $g$ and $h_{\omega}$, then \eqref{costo to go intermediate recursion} equals
\begin{equation}\nonumber
    \prod_{j = k+1}^N g(\pi^j  |  \pi^{j+1:N} ) \sum_{\omega \in \Omega^{k+1:N} } h_\omega(\Delta_{k+1}) \int_\nabla P(\pi^k | z) \frac{P(\omega  |  z)}{\psf(\omega)} p_{\Delta_{k}}(x, \d z).
\end{equation} 
Proposition \ref{prop:prod-cond-SM}  now gives
\begin{equation}\nonumber
\begin{aligned}
    &\, \prod_{j = k+1}^N g(\pi^j  |  \pi^{j+1:N} ) \sum_{ \substack{ \omega \in \Omega^{k+1:N} \\ \mu \in \coag(\pi^k, \omega)} } h_\omega(\Delta_{k+1}) \H(\pi^k, \omega  |  \mu) \frac{\psf(\mu)}{\psf(\omega)} \int_\nabla   \frac{P(\mu  |  z)}{\psf(\mu)} p_{\Delta_{k}}(x, \d z) 
\end{aligned}
\end{equation} 
and using the duality (recall Section \ref{SMsubsec:dual}), we find 
\begin{equation}\nonumber
    \prod_{j = k+1}^N g(\pi^j  |  \pi^{j+1:N} )  \sum_{ \substack{ \omega \in \Omega^{k+1:N} \\ \mu \in \coag(\pi^k, \omega)} } h_\omega(\Delta_{k+1}) \Pr(\mu, \pi^k  |  \omega) \sum_{\eta \sp \mu } \dualtr_{\mu, \eta}(\Delta_{k}) \frac{P(\eta  |  x)}{\psf(\eta)}.
\end{equation} 
The sums in the last display can be written
\begin{align}
    \sum_{\eta \in \Omega^{k:N} } \frac{P(\eta  |  x)}{\psf(\eta)} \sum_{ \mu \in \coag(\pi^k, \Omega^{k+1:N}) } \dualtr_{\mu, \eta}(\Delta_{k}) \sum_{\omega \in \Omega^{k+1:N} } h_\omega(\Delta_{k+1}) \Pr(\mu, \pi^k  |  \omega).
\end{align}
Now multiply and divide by
\begin{equation}\nonumber
g(\pi^k  |  \pi^{k+1:N} ):=\sum_{\omega \in \Omega^{k+1:N} } h_\omega(\Delta_{k+1}) \psf(\pi^k  |  \omega) 
\end{equation} 
and denote
\begin{equation}\label{small h of omega}
    h_\eta(\Delta_k) := \sum_{ \mu \in \coag(\pi^k, \Omega^{k+1:N}) } \left[ \sum_{\omega \in \Omega^{k+1:N} } \frac{h_\omega(\Delta_{k+1}) \psf^{\omega}(\pi^{k}) }{ g(\pi^k  |  \pi^{k+1:N} ) } \Pr(\mu  |  \omega, \pi^k) \right] \dualtr_{\mu, \eta}(\Delta_{k}),
\end{equation}
which is the probability the dual process with initial distribution on $\coag(\pi^k, \Omega^{k+1:N})$ given by the expression in square brackets
is in state $\eta \in \Omega^{k:N}$ after a time interval  equal to $\Delta_{k}$.
The induction is completed as we have found that 
\begin{equation}
     \Pr(\pi^{k:N}  |  X(t_{k-1}) = x) = \prod_{j = k}^N g(\pi^j  |  \pi^{j+1:N} ) \sum_{\omega \in \Omega^{k:N} } h_\omega(\Delta_k) \frac{P(\omega  |  x)}{\psf(\omega)}.
\end{equation}
\qed

\emph{Proof of Theorem \ref{thm:smoothing}}.
Consider first the following decomposition of the marginal smoothing distribution $\nu_{k|0:N}$,
\begin{equation}\label{precomputation on smoothing}
\begin{aligned}
    \nu_{k | 0:N}(A)
    =&\, \Pr(X(t_k) \in A  |  \pi^{0:N} ) \\
    \propto&\,  \int_A \Pr(X(t_k) \in \d x  |  \pi^{0:k}) \Pr(\pi^{k+1:N}  |  X(t_k) = x) \\
    =&\, \int_A \nu_{k | 0:k}(\d x) \Pr(\pi^{k+1:N}  |  X(t_k) = x).
\end{aligned}
\end{equation} 
where the random partitions $\Pi^{0:N}$ are not included in $\pr$ just to ease the notation.
Hence, up to proportionality, $\nu_{k | 0:N}$ factorizes into  $\nu_{k|0:k}$ and the likelihood of partitions collected after time $t_{k}$, the latter given in Proposition \ref{thm:cost to go}.
As  the first factor in \eqref{precomputation on smoothing} is the filtering distribution $\nu_{k|0:k}$ in the statement, we now have
\begin{equation}
    \nu_{k | 0:N}(\d x) \propto \sum_{ \substack{ \omega \in \Omega^{k:N} \\ \lambda \in \Lambda_{0:k} } } w_\lambda h_\omega(\Delta_{k+1}) \PD^\lambda(\d x) \frac{P(\omega  |  x)}{\psf(\omega)}.
\end{equation}
Writing $\PD^\lambda(\d x)$ as in \eqref{conditional PSF} and applying Proposition \ref{prop:prod-cond-SM} we get
\begin{align}
    \nu_{k | 0:N}(\d x)
    \propto&\, \sum_{ \substack{ \omega \in \Omega^{k:N} \\ \lambda \in \Lambda_{0:k} } }  w_\lambda h_\omega(\Delta_{k+1}) \frac{P(\lambda  |  x) P(\omega  |  x) }{ \psf(\lambda) \psf(\omega)  } \PD(\d x) \\
    =&\, \sum_{ \substack{ \omega \in \Omega^{k:N} \\ \lambda \in \Lambda_{0:k}  } } w_\lambda \frac{h_\omega(\Delta_{k+1})}{\psf(\omega)} \sum_{\mu \in \coag(\lambda, \omega)} \H(\lambda, \omega  |  \mu) \frac{P(\mu  |  x)}{\psf(\lambda)} \PD(\d x) \\
    =&\, \sum_{ \substack{ \omega \in \Omega^{k:N} \\ \lambda \in \Lambda_{0:k}} } w_\lambda \frac{h_\omega(\Delta_{k+1})}{\psf(\omega)} \sum_{\mu \in \coag(\lambda, \omega)} \H(\lambda, \omega  |  \mu) \frac{\psf(\mu)}{\psf(\lambda)} \PD^\mu(\d x) \\
    =&\, \sum_{ \substack{ \omega \in \Omega^{k:N} \\ \lambda \in \Lambda_{0:k}  } } w_\lambda \frac{h_\omega(\Delta_{k+1})}{\psf(\omega)} \sum_{\mu \in \coag(\lambda, \omega)} \Pr(\mu, \omega  |  \lambda) \PD^\mu(\d x).
\end{align}
where $\Pr(\mu, \omega  |  \lambda)$ is a shorthand notation for the conditional distribution in Proposition \ref{prop:joint-crp}.
Here the sums are finite, so inverting them leads to the mixture
\begin{align}
\sum_{ \mu \in \coag(\Lambda^{0:k}, \ \Omega^{k:N} ) } \left[ \sum_{ \substack{ \omega \in \Omega^{k:N} \\ \lambda \in \Lambda_{0:k}  } } \frac{ w_\lambda \psf^{\lambda}(\omega)  h_\omega(\Delta_{k+1}) }{\psf(\omega)}  \Pr(\mu | \lambda, \omega) \right] \PD^\mu(\d x),
\end{align}
which concludes the proof upon noting that
$$\Pr(\lambda, \omega) = \psf(\lambda) \psf^{\lambda}(\omega) = \psf(\omega) \psf^{\omega}(\lambda),$$
which follows from \eqref{conditional PSF}.
\qed

\emph{Proof of Corollary \ref{cor:interpolation}}.
    Following the same steps in the proof of Theorem \ref{thm:smoothing},
    \begin{equation} \nonumber
    \begin{aligned}
        \Pr(X(t_k + \delta) \in \d x  |  \pi^{0:N})
        &\propto \Pr(X(t_k + \delta) \in \d x  |  \pi^{0:k}) \Pr(\pi^{k+1:N}  |  X(t_k + \delta) = x) \\
        &= \varphi_\delta(\nu_{0:k})(\d x) \Pr(\pi^{k+1:N}  |  X(t_k + \delta) = x).
    \end{aligned}
    \end{equation}
    where the random partitions $\Pi^{0:N}$ are omitted in $\Pr$ just to ease the notation.
    As already shown in Theorem \ref{thm:propagation}, the propagation of $\nu_{k|0:k}$ is
    \begin{equation} \nonumber
        \varphi_\delta(\nu_{k|0:k}) = \sum_{\omega \in L(\Lambda_{0:k})} v_\omega(\delta) \PD^\omega.
    \end{equation}
    In order to find the likelihood $\Pr(\pi^{k+1:N}  |  X(t_k + \delta) = x)$ one can proceed as in the proof of Proposition \ref{thm:cost to go}, but replacing $\dualtr_{\mu, \eta}(\Delta_{k+1})$ with $\dualtr_{\mu, \eta}(\Delta_{k+1} - \delta)$.\qed

\emph{Proof of Corollary \ref{cor: forecasting}.}
The result is immediate upon writing the left-hand side of \eqref{eq: forecasting} as
\begin{equation}\nonumber
 \int_{\nabla} P(\pi  |  x) \Pr(X(t) \in \d x  |  \pi^{0:N})
 =\sum_{\lambda \in \Lambda_{t}} w_\lambda  \int_{\nabla} P(\pi  |  x) \PD^\lambda(\d x)
\end{equation} 
and using \eqref{conditional PSF}.\qed


\section{Implementation}\label{SM:alg}

\spacingset{.8}
\subsection{Pseudocodes}

\begin{algorithm}[H]
    \caption{Sampling $\varepsilon$-$\PD$}
        \small
    \label{alg: truncated PD}
    \KwIn{parameters $\alpha, \theta$ ; truncation error $\varepsilon \in (0, 1)$}
    
    $r \gets 1$ \\
    
    $i \gets 0$ \\
    
    \While{$r > \varepsilon$}
    {
    $i \gets i + 1$ \\
    
    Draw $V_i \sim \text{Beta}(1 - \alpha, \theta + i\alpha)$ \\
    
    $X_i \gets V_i \prod_{j < i} (1-V_j)$ \\

    $r \gets r - X_i$ \\
    }

    \Return $(X_1, \dots, X_i)$ in decreasing order (to be normalized if needed)
\end{algorithm}

\bigskip

\begin{algorithm}[H]
\caption{Sampling $\varepsilon$-$\PD^\pi$}
\small
\label{alg: conditional PD} 
\KwIn{partition $\pi \in \P$, truncation error $\varepsilon \in (0, 1)$}
Draw $W \sim \mathrm{Beta}(\theta + \len \alpha, |\pi| - \len \alpha)$ \\
Draw $V^{(1)}\sim\mathrm{Dir}(\pi_1 - \alpha, \dots, \pi_{\len} - \alpha)$ \\
Draw $V^{(2)} \sim \pd_{\alpha, \theta + \alpha \len}$ as in Algorithm \ref{alg: truncated PD} with truncation error $\varepsilon / (1-W)$ \\
\Return $((1-W) V^{(1)}_{1},\ldots,(1-W) V^{(1)}_{\len},W V^{(2)}_{1},W V^{(2)}_{2},\ldots)$ in decreasing order (to be normalized if needed)
\end{algorithm}

\bigskip

\begin{algorithm}[H]
\caption{Propagation -- Sampling $X(t_{k})$ given $\pi^{0:k-1}$ (exact)}
\small
\label{alg: signal prediction} 
        \KwIn{Active nodes $\Lambda\subset \P$ and weights $\{w_\lambda,\lambda\in \Lambda\}$ from $\nu_{k-1|0:k-1}$; sample size $M$}
\For{$\lambda \in \Lambda$}{
    \For{$\omega \sp \lambda$}{
        Initialization: $v_{\omega} \gets 0$ if $v_{\omega}$ is NULL \\
        $v_{\omega} \gets v_{\omega} + w_\lambda \dualtr_{\lambda, \omega}(\Delta_{k}) $  \\
}
}
Normalize $\{v_{\omega}, \omega \in L(\Lambda)\}$\\
Draw $\omega_m \sim v_{\omega}$ and $X_m \sim \PD^{\omega_m}$ using Algorithm \ref{alg: conditional PD}, for $m = 1, \dots, M$ \\
\Return $\{v_{\omega}, \omega \in L(\Lambda)\}$ and $\{X_m, m = 1, \dots, M\}$
\end{algorithm}

\bigskip

\begin{algorithm}[H]
\caption{Propagation -- Sampling $X(t_{k})$ given $\pi^{0:k-1}$ (Gillespie)}
\small
\label{alg: signal prediction gillespie} 
        \KwIn{Active nodes $\Lambda\subset \P$ and weights $\{w_\lambda,\lambda\in \Lambda\}$ from $\nu_{k-1|0:k-1}$; sample size $M$}
\For{$m = 1, \dots, M$}{
Draw $\eta \sim w_\eta$ \\
Draw $E \sim \text{Exp}(|\eta|(|\eta| + \theta - 1) / 2)$  \\
$\Delta \gets E$ \\
\While{$\Delta < \Delta_k$}{
Delete uniformly at random one square from the Young diagram of $\eta$ \\
Draw $E \sim \text{Exp}(|\eta|(|\eta| + \theta - 1) / 2)$  \\
$\Delta \gets \Delta + E$
}
Initialization: $v_{\eta} \gets 0$ if $v_{\eta}$ is NULL\\
$v_{\eta} \gets v_{\eta} + 1 $\\
Draw $X_m \sim \PD^\eta$ using Algorithm \ref{alg: conditional PD}\\
}
\Return $\{v_{\eta}, \eta \in L(\lambda)\}$ normalized and $\{X_m, m = 1, \dots, M\}$\\
\end{algorithm}

\bigskip

\begin{algorithm}[H]
\caption{Update -- Sampling $X(t_{k})$ given $\pi^{0:k}$}
\small
\label{alg: updated prediction} 
    \KwIn{$\{w_\lambda,\lambda\in \Lambda\}$ from $\nu_{k|0:k-1}$; partition $\pi^k$; sample size $M$}
    $\coag(\Lambda,\pi^k) \gets \varnothing$ \\
\For{$\lambda \in \Lambda $}{
    \For{$\mu \in \coag(\lambda, \pi^k)$}{
        $\coag(\Lambda, \pi^k) \gets \coag(\Lambda, \pi^k) \cup \mu $  \\
        Initialization: $v_\mu \gets 0$ if $v_\mu$ is NULL \\
        $v_\mu \gets v_\mu + \frac{w_\lambda}{\psf(\lambda)} \H(\lambda, \pi^k | \mu) \psf(\mu)$  \\
}
}
Normalize $\{v_\mu, \mu \in \coag(\Lambda, \pi^k) \}$\\
Draw $\mu_m \sim v_\mu$ and $X_m \sim \PD^{\mu_m}$ using Algorithm \ref{alg: conditional PD}, for $m = 1, \dots, M$\\
\Return $\{v_\mu, \mu \in \coag(\Lambda, \pi^k) \}$ and $\{X_m, m = 1, \dots, M\}$\\
\end{algorithm}

\bigskip

\begin{algorithm}[H]
\caption{Filter}
\small
\label{alg: generic filtering} 
\KwIn{Partitions $\pi^{0:N}$}
\textbf{Update:} Let $\nu_{0|0} = \PD^{\pi^0}$ (and sample from it), i.e. initialise with $\Lambda = \{ \pi^0\}$ and $v_{\pi^0} = 1$ \\
\For{ $k = 1, \dots, N$ }{
\textbf{Propagation:} Evaluate (and sample from, if needed) $\nu_{k|0:k-1}$ using Algorithm \ref{alg: signal prediction}\\
\textbf{Update:} Evaluate (and sample from) $\nu_{k|0:k}$ using Algorithm \ref{alg: updated prediction} \\
}
\end{algorithm}

\bigskip

\begin{algorithm}[H]
\caption{Smoother -- Sampling $X(t_{k})$ given $\pi^{0:N}$}
\small
\label{alg: smoothing} 
\KwIn{Partitions $\pi^{0:N}$; sample size $M$}
Evaluate $\nu_{k | 0:k}$ using Algorithm \ref{alg: generic filtering}, i.e. compute $\{ w_\lambda, \lambda \in \Lambda_{0:k} \}$ \\
Evaluate $\nu_{k|k+1:N}$, i.e. propagate the result of Algorithm \ref{alg: generic filtering} with $\pi^{N}, \pi^{N-1}, \dots, \pi^{k+1}$ given as input in this order, i.e. compute $\{h_\omega, \omega \in \Omega^{k:N} \}$ \\
$\coag(\Lambda_{0:k}, \Omega^{k:N}) \gets \varnothing$ \\
\For{$\lambda \in \Lambda_{0:k}, \omega \in \Omega^{k:N}$}{
    \For{$\mu \in \coag(\lambda, \omega)$}{
        $\coag(\Lambda_{0:k}, \Omega^{k:N}) \gets \coag(\Lambda^{0:k}, \Omega^{k:N}) \cup \mu$  \\
        Initialization: $s_\mu \gets 0$ if $s_\mu$ is NULL \\
        $s_\mu \gets s_\mu + \frac{v_\lambda}{\psf(\lambda)} \frac{h_\omega}{\psf(\omega)} \H(\lambda, \omega | \mu) \psf(\mu)$ \\
    }
}
Normalize $\{ s_\mu, \mu \in \coag(\Lambda_{0:k}, \Omega^{k:N})\}$ \\
Draw $\mu_m \sim s_\mu$ and $X_m \sim \PD^{\mu_m}$ using Algorithm \ref{alg: conditional PD}, for $m = 1, \dots, M$\\
\Return $\{ s_\mu, \mu \in \coag(\Lambda_{0:k}, \Omega^{k:N})\}$ and $\{X_m, m = 1, \dots, M\}$
\end{algorithm}

\bigskip

\begin{algorithm}[H]
\caption{Sampling further partitions given $ \pi^{0:N}$ }
\small
\label{alg: forecasting} 
\KwIn{Partitions $\pi^{0:N}$; weights $\{w_\lambda,\lambda\in \Lambda\}$ from $\Pr(X(t) \in \d x  |  \pi^{0:N})$; partition size $m$}
Draw $\lambda \sim w_{\lambda}$\\
Simulate the next $m$ steps of the $\CRP$ started from $\lambda$, i.e. sample $\Pi_{1:|\lambda|+m} | \Pi_{1:|\lambda|} = \lambda $\\ 
\Return the partition $\Pi_{|\lambda|+1:|\lambda|+m}$ induced by the last $m$ customers
\end{algorithm}

\spacingset{1.1}


\subsection{Weights of filtering and smoothing distributions}\label{SM:mixw}

We collect here additional details on the mixture weights for filtering and smoothing.

\begin{figure}[ht]
    \centering
    \includegraphics[width=.75\linewidth]{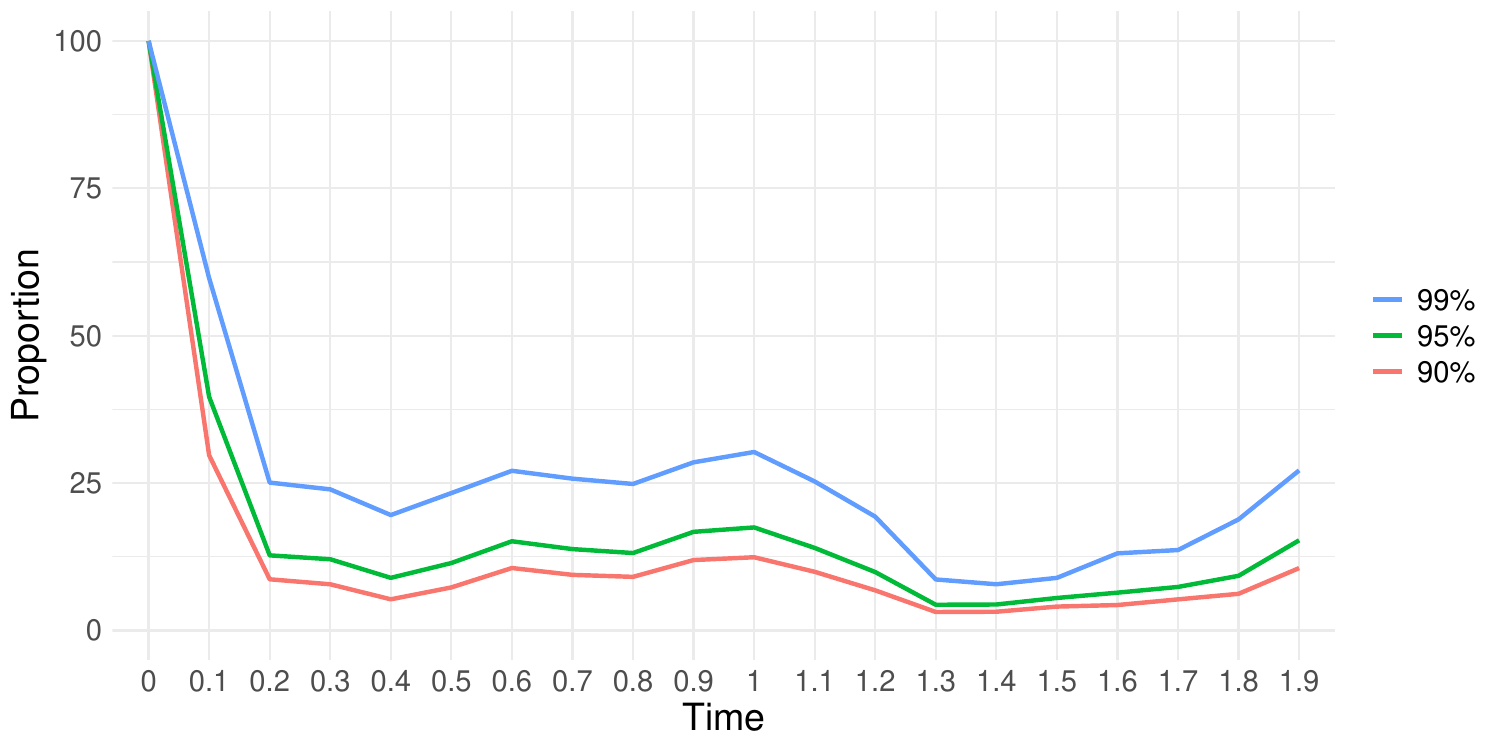}
    \caption{\scriptsize Proportion of filtering mixture components needed to reach 90\%, 95\%, and 99\% cumulative probability.}
    \label{fig:mixture cardinality}
\end{figure}

\paragraph{Filtering.}
Empirically, most probability mass concentrates on a small subset of mixture components, which supports pruning. Figure~\ref{fig:mixture cardinality} illustrates the proportion of mixture components required to reach different cumulative thresholds. 
Simulation results (with $T=19$, $\Delta_k=0.1$, $\alpha=0.1$, $\theta=1.5$, partitions of size $10$, and $10^5$ Gillespie particles) show that pruning away the least-weighted components retains nearly all the mass. Cf.~Figure~\ref{fig:time_vs_pruning_well_specified} for a performance comparison of pruning strategies.

\paragraph{Smoothing.}
The smoothed law $\nu_{k|0:N}$ from Theorem~\ref{thm:smoothing} combines information from past data $\pi^{0:k}$ and future data $\pi^{k+1:N}$. For any measurable $A\subset\nabla$,
\[
\nu_{k|0:N}(A)
=\int_A \nu_{k|0:k}(\d x)\,
\Pr\!\big(\Pi^{k+1}=\pi^{k+1},\dots,\Pi^N=\pi^N |  X(t_k)=x\big),
\]
(cf.~\eqref{precomputation on smoothing}), which separates the forward contribution $\nu_{k|0:k}$ (Theorem~\ref{thm:update}) from the backward contribution (the cost-to-go likelihood) conditional on the state $x$.

In practice, evaluating $\Coag(\lambda,\omega)$ and its probabilities is most convenient via Proposition~\ref{prop:joint-crp}, using
\[
\Pr(\Pi_{1:|\lambda|}=\lambda,\Pi_{|\lambda|+1:|\lambda|+|\omega|}=\omega)
= \sum_{\mu\in \coag(\lambda,\omega)} \H(\lambda,\omega |  \mu)\,\psf(\mu),
\]
and
\[
\Pr(\Pi_{1:|\lambda|+|\omega|}=\mu  |  \Pi_{1:|\lambda|}=\lambda,\Pi_{|\lambda|+1:|\lambda|+|\omega|}=\omega)
=\frac{\H(\lambda,\omega |  \mu)\,\psf(\mu)}
{\sum_{\mu'\in \coag(\lambda,\omega)} \H(\lambda,\omega |  \mu')\,\psf(\mu')}.
\]
Consequently, the smoothing weights can be computed in the equivalent, implementation-friendly form
\[
s_\mu(\Delta_{k+1})\ \propto\
\sum_{\lambda\in \Lambda_{0:k}} \sum_{\omega\in \Omega^{k:N}}
\frac{v_\lambda}{\psf(\lambda)}\,
\frac{h_\omega(\Delta_{k+1})}{\psf(\omega)}\,
\H(\lambda,\omega |  \mu)\,\psf(\mu),
\]
which is what Algorithm~\ref{alg: smoothing} evaluates.


\subsection{Further figures}\label{SM:param}

\begin{figure}[ht]
    \centering
    \includegraphics[width=\linewidth]{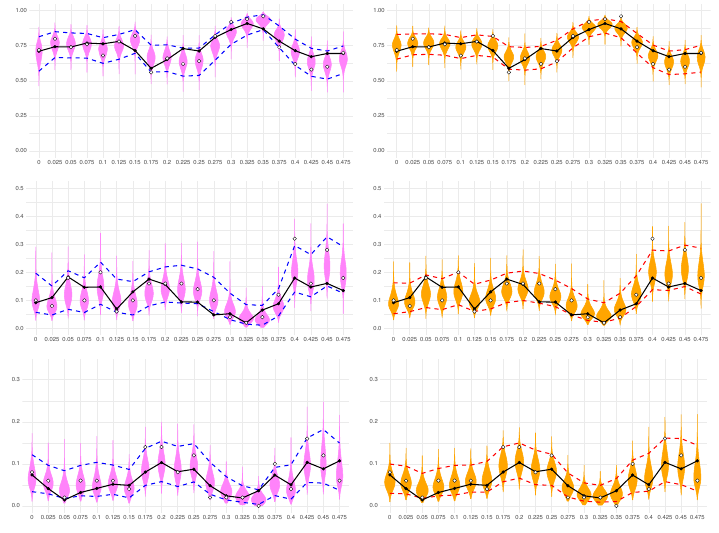}
    \caption{\scriptsize Filtering (left) and smoothing (right) posterior estimates for the three largest coordinates of the system. Violin plots with dashed intervals show posterior distributions and 95\% credible intervals; solid black lines denote the true values; diamonds indicate observed relative frequencies.}
    \label{fig:coord_filter_smooth_mle}
\end{figure}

Figure~\ref{fig:coord_filter_smooth_mle} displays filtering (left column) and smoothing (right column) estimates of the three largest system coordinates (top to bottom), as described in Sections~\ref{SMsec:prediction} and~\ref{SMsec:smoothing}. The plotted data (diamonds) are the observed relative frequencies at the corresponding ranks. For example, the second panel shows the relative frequency of the second-largest group. Although presented separately for visualization, these estimates are obtained jointly for the full $\simplexone$-valued vector, and the data are likewise used jointly to condition the model.

\begin{figure}[ht]
    \centering
    \includegraphics[width=\linewidth]{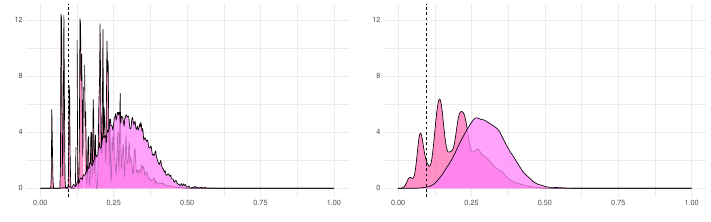}
    \caption{\scriptsize Marginal filtering densities at the fifth observation time, comparing the bootstrap particle filter (pink) with the dual filter (blue) under two kernel bandwidths (left: 0.001; right: 0.01). The vertical dashed line marks the observed heterozygosity.}
    \label{fig:marginal}
\end{figure}

In Section~\ref{sec: application} of the main text we noted that the bootstrap particle filter (BPF) can produce multimodal posterior densities, which is undesirable. This occurs, for instance, at the fifth observation time. 
Figure~\ref{fig:marginal} provides a magnified comparison of the marginal posterior densities at that time, contrasting the BPF with the duality-based filter for two kernel bandwidths (0.001 and 0.01). Despite identical bandwidths, the dual filter yields smoother posteriors. While the multimodality under the BPF could be mitigated by increasing the number of particles, this is computationally prohibitive in practice.


\spacingset{1.03}   
\setlength{\bibsep}{4pt}

\bibliography{bib_file}
\bibliographystyle{apalike}

\end{document}